\documentclass[10.5pt,letter,doc,natbib,donotrepeattitle,floatsintext]{apa6}

\usepackage[english]{babel}
\usepackage[utf8x]{inputenc}
\usepackage{amsthm,amsmath,bm,bbm}
\usepackage{amssymb,mathtools}
\usepackage{graphicx}
\usepackage[dvipsnames]{xcolor}
\usepackage[colorinlistoftodos]{todonotes}
\usepackage{setspace}
\usepackage[font=singlespacing]{caption}
\usepackage{tabulary}
\usepackage{multirow}
\usepackage{float}
\usepackage{hyperref}
\usepackage{array}
\usepackage{etoolbox}
\patchcmd{\maketitle}{\newpage}{}{}{}
\usepackage{tikz}
\usetikzlibrary{shapes,shadows,arrows,decorations.pathreplacing}

\usepackage{resizegather}
\usepackage{soul}

\usepackage{pdfpages}

\def\independent{\perp\!\!\!\perp}
\def\E{\text{E}}
\def\P{\text{P}}

\DeclareSymbolFont{rsfs}{U}{rsfs}{m}{n}
\DeclareSymbolFontAlphabet{\mathscrsfs}{rsfs}

\AtBeginDocument{%
 \abovedisplayskip=3pt plus 2pt minus 2pt
 \abovedisplayshortskip=0pt plus 2pt
 \belowdisplayskip=3pt plus 2pt minus 2pt
 \belowdisplayshortskip=3pt plus 2pt minus 2pt
}

\title{\Large Clarifying causal mediation analysis for the applied researcher:\\[.3em]Defining effects based on what we want to learn}

\shorttitle{Causal mediation analysis clarity}
\author{\small Trang Quynh Nguyen, Ian Schmid, Elizabeth A. Stuart}
\affiliation{\small Johns Hopkins Bloomberg School of Public Health\\~\\Manuscript accepted by Psychological Methods 2020-04-27.\\Published version doi: 10.1037/met0000299.}
\authornote{This work was supported by Grant NIMH R01MH115487 from the National Institute of Mental Health. The authors appreciate thoughtful feedback from three anonymous reviewers and from Daniel Malinsky and Fan Li, which helped improve the content and clarity of the article. Correspondance should be addressed to trang.nguyen@jhu.edu.}

\abstract{The incorporation of causal inference in mediation analysis has led to theoretical and methodological advancements -- effect definitions with causal interpretation, clarification of assumptions required for effect identification, and an expanding array of options for effect estimation. However, the literature on these results is fast-growing and complex, which may be confusing to researchers unfamiliar with causal inference or unfamiliar with mediation. The goal of this paper is to help ease the understanding and adoption of causal mediation analysis. It starts by highlighting a key difference between the causal inference and traditional approaches to mediation analysis and making a case for the need for explicit causal thinking and the causal inference approach in mediation analysis. It then explains in as-plain-as-possible language existing effect types, paying special attention to motivating these effects with different types of research questions, and using concrete examples for illustration. This presentation differentiates two perspectives (or purposes of analysis): the explanatory perspective (aiming to explain the total effect) and the interventional perspective (asking questions about hypothetical interventions on the exposure and mediator, or hypothetically modified exposures). For the latter perspective, the paper proposes tapping into a general class of interventional effects that contains as special cases most of the usual effect types -- interventional direct and indirect effects, controlled direct effects and also a generalized interventional direct effect type, as well as the total effect and overall effect. This general class allows flexible effect definitions which better match many research questions than the standard interventional direct and indirect effects.}

\keywords{mediation, causal mediation, effect definition, identification, assumptions, interventional effects, natural effects}

\begin{document}
\pagenumbering{gobble}
\maketitle
\pagenumbering{arabic}

\begin{figure*}[h!]
\caption{An increasing trend in the popularity of mediation analysis in scholarly research}\label{fig:MediationAnalysisTrends}
\centering
\includegraphics[width=.75\linewidth]{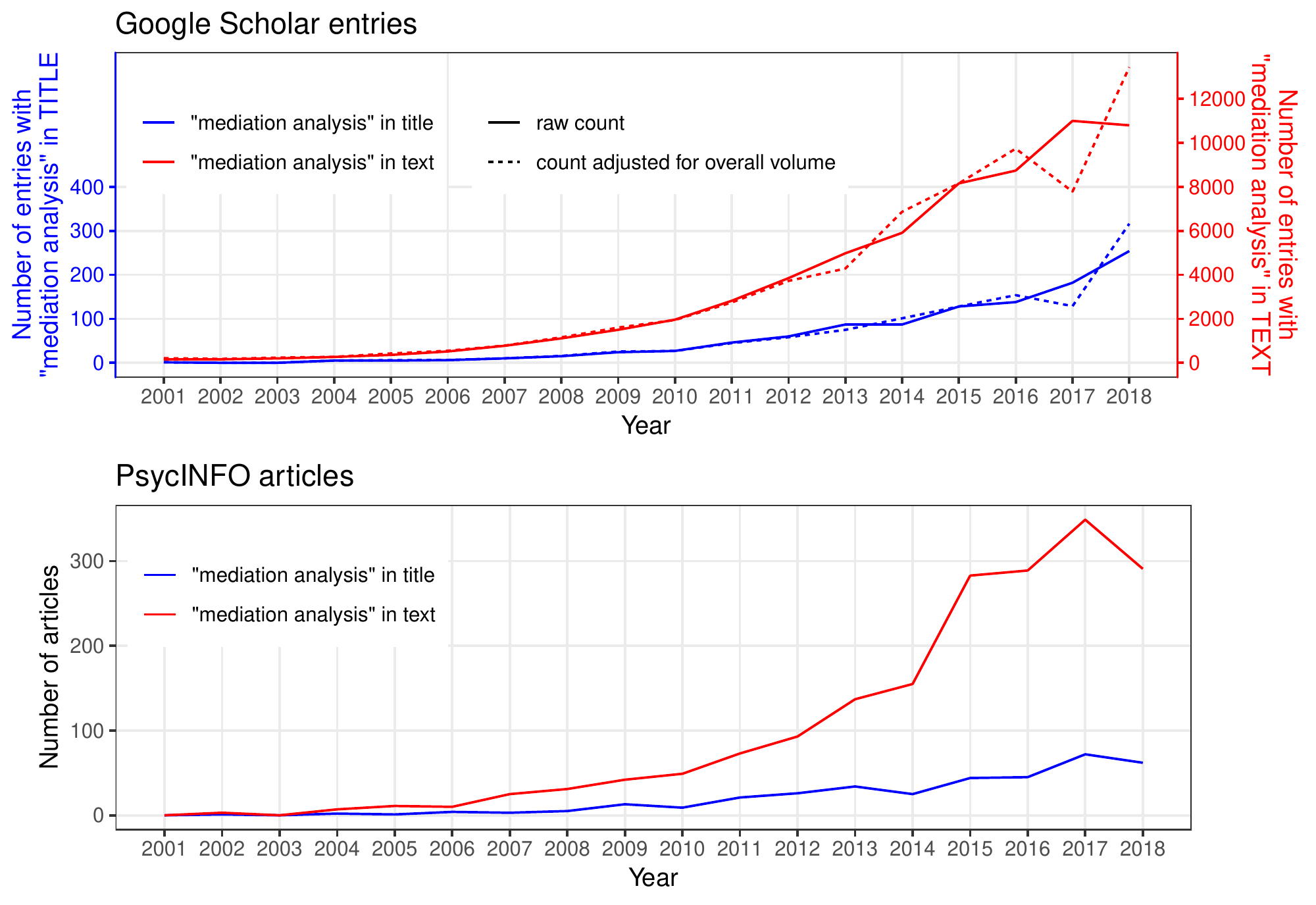}
\caption*{\footnotesize The raw counts in the top panel are counts reported by Google Scholar on two searches for articles (excluding patents and citations) with ``mediation analysis'' in the title, and for those with the same phrase anywhere in the text; the adjusted counts are adjusted for the fact that the volume of all Google Scholar entries varies in size from year to year, using 2015 as the standard year. In the bottom panel, the counts are reported by PsycINFO on the same two searches. These searches were conducted on 20/12/2018.}
\end{figure*}

\noindent Mediation analysis is becoming more popular. Fig. \ref{fig:MediationAnalysisTrends} shows that both the number of entries in Google Scholar and the number of peer-reviewed articles in PsycINFO that have ``mediation analysis'' in the title or text have been growing exponentially. While researchers may have long been interested in causal processes, mediation analysis has become more accessible after several decades of accumulation of theory, methods and computing tools. In addition, the investigation of mediators (mechanisms) of intervention effects is now encouraged, or even required, by some research funding agencies, e.g., the National Institute of Mental Health \citep{RFA}. With these push and pull forces, the increased interest in mediation analysis will likely continue, and mediation analyses may exert increasing influence on policy and practice.

Mediation analysis is not new -- the idea dates back to at least as early as \cite{Wright1934}, and the seminal \cite{Baron1986} paper that popularized mediation analysis in the social sciences was published more than 30 years ago. Yet the methods of mediation analysis are still an active area of research. A major advancement in more recent years is the incorporation of the \textit{causal inference} approach. This has led to \mbox{(1) formulation} of effect definitions that are more general than those from the prior mainstream (hereafter \textit{traditional}) approach and that have causal interpretations, and \mbox{(2) clarification} of  the assumptions required for such effects to be identified from data, allowing researchers to scrutinize these assumptions based on substantive knowledge; and has opened up \mbox{(3) a} range of relevant estimation methods. However, the methodological literature on causal mediation analysis is fast-growing and complex, which may be confusing to applied researchers and methodologists alike -- both those unfamiliar with causal inference and those familiar with causal inference but not specifically with mediation.

Our goal with this paper is to help ease the understanding and adoption of causal mediation analysis. To start, the paper highlights a key distinction between the causal inference and traditional approaches to mediation analysis, and makes a case for explicit causal thinking and the causal inference approach. The bulk of the paper then focuses on the first order of business in causal mediation analysis, defining the target causal effect(s). This first step is important because the effect(s) targeted by an analysis should reflect the research question (what the researchers want to learn), and clarity about this helps the researchers appropriately communicate the results of the analysis (what they have learned). We explain, in one place and in as-plain-as-possible language, several existing effect types: controlled direct effects, natural direct and indirect effects (Robins \& Greenland, \citeyear{Robins1992}; \citeauthor{Pearl2001}, \citeyear{Pearl2001}), and interventional direct and indirect effects \citep{Didelez2006,Lok2016,VanderWeele2014a}. Using concrete examples, we illustrate the types of research questions these effects are fit to answer. This discussion differentiates two general perspectives (or purposes of analysis): the \textit{explanatory} perspective, aiming to explain the total causal effect; and the \textit{interventional} perspective, asking questions about hypothetical interventions on the exposure and mediator, or hypothetically modified exposures. We also argue that, if adopting the latter perspective, in many cases we should tap into a more general class of interventional effects rather than restricting to the standard interventional direct and indirect effects.

\section{The key difference between the causal inference and traditional approaches}

\noindent A typical mediation analysis seeks to understand whether, and to what degree, the effect of an exposure $A$ on an outcome $Y$ involves changing an intermediate variable $M$. How should such an analysis be done? The answer depends on what is meant by the effect of $A$ on $Y$ through $M$ (the \textit{indirect} effect). This differs between the two approaches.

\begin{figure*}[ht!]
    \caption{Traditional approach: effects defined as (functions of) regression coefficients}\label{fig:traditional}
    \begin{center}
    \begin{tikzpicture}[
    obs/.style={rectangle, minimum size=5mm}
    ]
        \node[obs]   (C)                            {$C$};
        \node[obs]   (A)  [right=of C, xshift=+3mm] {$A$};
        \node[obs]   (M)  [above right=of A, xshift=+3mm] {$M$};
        \node[obs]   (Y)  [below right=of M, xshift=+3mm] {$Y$};
        \node[obs] [right=of A, xshift=-8mm, yshift=+9mm] {$a$};
        \node[obs] [left=of Y, xshift=+7mm, yshift=+10mm] {$b$};
        \node[obs] [right=of A, xshift=+4mm, yshift=2mm] {$c'$};
        \node[obs] [right=of Y, yshift=+7mm] {\footnotesize\begin{tabular}{l}$a$: coef of $A$ in model $M\sim A+C$\\$b,c'$: coefs of $M$ and $A$ in model $Y\sim A+M+C$\\ \\indirect effect $:=ab$\\direct effect $:=c'$\end{tabular}};
        \draw[->] (C) -- (A);
        \draw[->] (C) .. controls +(down:10mm) and +(down:10mm)  .. (Y);
        \draw[->] (C) -- (A);
        \draw[->] (C) .. controls +(up:10mm) and +(left:10mm)  .. (M);
        \draw[thick, ->] (A) -- (M);
        \draw[thick, ->] (A) -- (Y);
        \draw[thick, ->] (M) -- (Y);
    \end{tikzpicture}
    \end{center}
\end{figure*}
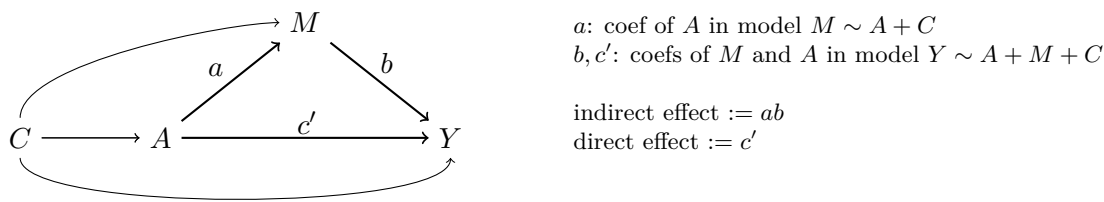

The traditional approach uses a model-based definition. It assumes two parametric models: (1) a model for the mediator $M$ with $A$ and covariates $C$ as predictors; and (2) a model for the outcome $Y$ with $A,M,C$ as predictors.%
\footnote{\cite{Baron1986} used linear models for a continuous mediator and a continuous outcome. Subsequent work added model options for a wide range of situations, such as noncontinuous mediator or outcome, multiple simultaneous or sequential mediators, multi-level mediator and/or outcome, conditional effects, etc. -- see comprehensive texts \cite{MacKinnon2008} and \cite{Hayes2018}.}\textsuperscript{,}%
\footnote{Many published analyses actually do not include covariates. This may be related to the unfortunate fact that the influential \cite{Baron1986} paper leaves out covariates. Its lesser known predecessor, the \cite{Judd1981} paper, however, devotes one whole section to bias due to excluding covariates.}
The \textit{indirect effect} is defined as the product of the coefficient of $A$ in the mediator model and the coefficient of $M$ in the outcome model. The idea is that these coefficients represent the effect of $A$ on $M$ and the effect of $M$ on $Y$, so their combination should represent the effect of $A$ on $Y$ through $M$. The \textit{direct effect} is defined to be the coefficient of $A$ in the outcome model.\footnote{If both models are linear, the product of coefficients is equal to the difference between the coefficient of $A$ in an outcome model with $A,C$ as predictors (leaving out $M$) and the coefficient of $A$ in the outcome model with $A,M,C$. The traditional mediation literature thus talks about the \textit{product} and \textit{difference} methods.} A triangle diagram is often used to depict these effect definitions (\citeauthor{Baron1986}, \citeyear{Baron1986}, page 1176; \citeauthor{MacKinnon2008}, \citeyear{MacKinnon2008}, page 49; \citeauthor{Hayes2018}, \citeyear{Hayes2018}, page 83), which we show in Fig. \ref{fig:traditional} with a small modification to represent covariates. The key point is, the indirect and direct effects here are mathematical objects that do not exist without the model; there is no separation of the definition of an effect and its estimation method. 

In contrast, the causal inference approach separates the definition of an effect that we researchers want to estimate from how we may estimate it.%
\footnote{We refer specifically to effects here, but this separation has broader relevance. In general, what we want to estimate is called the \textit{estimand}, and a method/procedure we use to estimate it is called an \textit{estimator}.} Effects are defined in a model-free manner, based on reasoning about what fits the notion of a causal effect. A helpful and popular framework for this purpose is the \textit{potential outcomes} (aka \textit{counterfactual}) framework (\citeauthor{Splawa-Neyman1923}, \citeyear{Splawa-Neyman1923}; \citeauthor{Rubin1974}, \citeyear{Rubin1974}; \citeauthor{Holland1986}, \citeyear{Holland1986}), in which a causal effect is defined as a contrast between potential outcomes under two different conditions, for the same individual, or the same group. The total effect for an individual (or for a group) is often defined as the difference between (i) the outcome the individual would have (or the average outcome the group would have) \textit{if exposed} to an exposure of interest and (ii) the outcome the same individual would have (or the average outcome the same group would have) \textit{if unexposed}. The total effect is thus a contrast of two conditions defined by values of the exposure. \cite{Robins1992} and \cite{Pearl2001} extended this reasoning to define an indirect (or a direct) effect as a contrast of two conditions defined by values of the exposure-mediator combination. How these conditions are formulated determines the type of direct and indirect (hereafter ``(in)direct'') effects; there are several types of these effects. The first task for a researcher is to determine which effects best match their research question.

After \textit{effect definition}, the next step in the causal inference approach is \textit{effect identification}, that is, determining whether the causal effect of interest can be learned from data (aka, is \textit{identified}). Methodologists have worked out the assumptions required for identification of each effect type; the task of the substantive researcher is to judge whether such assumptions are plausible with their own data. Identification gives us the license to then estimate the causal effect, and it is only at this \textit{effect estimation} step that modeling questions (e.g., whether to fit a linear model for the mediator and a logistic model for the outcome, or do something else) come into the picture. This separation of the three steps -- effect definition, identification, and estimation -- is different from the traditional approach.

\section{The need for explicit causal thinking in mediation analysis}

\noindent We argue that the practice of mediation analysis would benefit from adopting explicit causal thinking. First, mediation analysis is, unavoidably, about causal effects. This is not the case with regular regression analysis, where the target may be either causal effects or conditional associations. In mediation analysis, the arrows drawn from $A$ to $M$ and $Y$ and from $M$ to $Y$ imply a conception that these variables affect one another in the directions depicted. The influence of $A$ on $Y$ through $M$ and the influence of $A$ on $Y$ through other ways -- referred to in the research question -- are causal effects. Therefore, a mediation analysis, whether using a traditional or causal inference method, is an analysis of causal effects.  In addition, results of mediation analysis are generally interpreted in causal terms -- regardless of any caution the authors may put in the limitations section. Hence analysis should be done in a way that aims to justify such interpretation. Note that this point, that mediation is conceptually causal, has been made repeatedly by prominent scholars in mediation methodology \citep[e.g.,][]{Baron1986,MacKinnon2008,Preacher2015}.

Second, while mediation analysis is about causal effects, such effects are not intuitively obvious in the mediation setting. In the non-mediation setting, intuition serves us researchers well in the pursuit of causal effects. For example, suppose we are interested in the effectiveness of a short-term college preparation program as an intervention to improve college readiness in high school students. If a trial randomizes high-schoolers to receive either this intervention or no intervention, and subsequently measures their readiness for college, it is rather obvious that the difference in college readiness between the students who received and those who did not receive the intervention represents the effect of the intervention. We may make this comparison and attribution intuitively, without consciously considering why. The implicit reasoning here, were we asked to explain ourselves, is that we observe the outcome in both of the conditions (intervention and no intervention) we wish to compare, and the two groups in the two conditions are similar (due to randomization) so it is reasonable to simply compare their outcomes. 
In the mediation setting, on the other hand, for a direct (or indirect) effect, we no longer have two observable conditions to compare. Suppose we wish to know the effect of this intervention that is mediated by an intermediate variable, self-awareness. Extending the reasoning above, we might wish, for example, to compare college readiness in two conditions: (a) the intervention condition and (b) a (hypothetically) modified intervention condition where the intervention is forced not to influence self-awareness (thus the intervention's effect on college readiness through self-awareness is blocked); this difference would be a fine notion of an indirect effect.%
\footnote{Alternatively, we might wish to compare to the no intervention condition a hypothetical condition where the intervention is allowed to influence college readiness only through influencing self-awareness.}
However, condition (b) does not exist, thus cannot be observed. The challenge of the mediation setting is that we do not have observable contrasts that correspond to lay perceptions of causal effects. Here, adopting explicit causal thinking allows us to be more clear about what exactly are the causal effects we want to learn (the focus of the later part of this paper), and to figure out whether, and how, we can learn it.

But there lingers the question: What is the problem with simply using the product of coefficients, which seems so intuitive? A brief answer: In the simplest case where linear models are used, for the product of coefficients to match a causal indirect effect (in the sense of the outcome contrast between conditions (a) and (b) in the example above), the following conditions are generally required: (1) all identification assumptions hold (roughly speaking, covariates $C$ are pre-exposure and capture all $A$-$M$, $A$-$Y$ and $M$-$Y$ confounding); (2) the additive effect of $M$ on $Y$ does not depend on the value of $A$ (aka no $A$-$M$ interaction) or on the base value of $M$ (aka $Y$ is linear in $M$); (3) either the additive effect of $A$ on $M$ or the additive effect of $M$ on $Y$ is the same across all individuals, or these two effects are independent of each other (aka constant or independent additive effects); and (4) the correct forms of $C$ are used in both models and do not interact with $A$ and $M$. (Similar assumptions are discussed in \citeauthor{MacKinnon2008}, \citeyear{MacKinnon2008}, chapter 3.) These are strong assumptions. The fourth one may be relaxed to some extent, e.g., when evaluating conditional effects within levels of $C$, but not the other three. The no $A$-$M$ interaction assumption is overly restrictive, as there are cases where such an interaction is expected \citep{MacKinnon2019}. If nonlinear models are used, the product of coefficients generally does not match, and may not be on the same scale as, the causal effect of interest.

Technical details aside, a general point is that as model assumptions are unlikely to hold, one may want to minimize them, or at least have flexibility in their selection; thus it makes sense to define effects in a model-free manner to have clarity about what we are trying to learn and avoid having a moving target. The causal inference approach fits this bill. To be clear, adopting the causal inference approach generally does not remove the need for some model assumptions at the estimation step. But instead of defaulting to a set of models, with model-free effect definitions we can consider what models to use based on the data at hand and based on what prior knowledge we have or do not have about the causal structure. Such consideration likely results in adopting more flexible models that make fewer assumptions,%
\footnote{A simple example is that models with $A$-$M$ interaction are not off limits, since estimation does not rely on multiplying one pair of regression coefficients.}
or assumptions that are more appropriate to the specific situation.

To mediation analysis, causal inference is a new approach, and it takes time for new approaches to take hold. A certain degree of interest in the causal inference perspective is seen in recent works by several methodologists known for substantial contribution in the traditional approach -- these works include causal inference as a topic in mediation analysis methodological reviews \citep{MacKinnon2007,MacKinnon2013}, emphasize confounding as a validity threat in mediation analysis \citep{MacKinnon2013,MacKinnon2015}, highlight the need to accommodate exposure-mediator interaction \citep{MacKinnon2019}, and explain how to implement causal mediation analysis \citep{Muthen2011,Miocevic2018,Muthen2015a}. In the applied research literature, however, the uptake of causal mediation analysis is still limited. In an ongoing review of articles published in top psychology and psychiatry journals in 2013-2018 that include a mediation analysis, we found that less than 4\% used causal mediation analysis. In our experience working with researchers (students, postdocs and faculty) in a public health research institution, we observe that often the researcher's less than ideal starting point (e.g., unfamiliarity with causal inference language, or loose understanding of confounding and confounding control) combined with the complexity of the methodology (e.g., multiple effect definitions and identification assumptions, different settings requiring different methods, and the explosion of the methodological literature) hinders adoption of causal mediation analysis. We aim to help address these barriers, starting with the current paper.

\section{Orientation to the rest of the paper}

\noindent Before delving into the causal effects, it is important to point out that this paper does not address analyses in the research literature which are unfortunately also referred to as mediation analyses, but do not reflect a setting where it is plausible for $A$ to influence $M$ and $Y$ and/or for $M$ to influence $Y$, for example due to lack of temporal ordering of these variables. These are analyses of associations (not causal effects), and in our opinion it is more appropriate to refer to them as ``third variable analyses''. Third variable analyses require a separate discussion that is outside the scope of the current paper.

Another third variable setting not covered here is where there is temporal ordering but the intermediate variable of interest is defined based on only one exposure condition (e.g., adherence to the active treatment or attendance of intervention sessions). For this problem a principal stratification \citep{Frangakis2002} approach could be used -- see \citet{Imai2011b,Jo2009,Jo2012,Jo2011}; note that causal effects defined in this approach are different from those discussed in the current paper. In our current mediation analysis setting, the mediator is a variable that is defined and has the same meaning for both exposure conditions; the key idea is the exposure (relative to nonexposure) makes a difference in the mediator, and that results in an effect on the outcome.

\medskip

Back to the task at hand, of the three steps in causal mediation analysis, the remainder of this paper focuses on the first step -- effect definition. We start with the total effect,  showing how it is defined based on potential outcomes at the individual and population levels, laying down basic causal inference concepts. We also introduce the \textit{causal directed acyclic graph} (DAG), a helpful tool for visualizing effect definitions and assumptions. We then bring in the mediator, and define, and discuss the practical relevance of, several effect types, first the natural (in)direct effects, then the interventional (in)direct effects, then a broader class of interventional effects, and ending with controlled direct effects.

This somewhat unusual order of presenting these effect types -- e.g., controlled direct effects not appearing first and not preceding natural (in)direct effects as in most papers that include both these effect types -- results from our attention to having the effects motivated by questions of potential practical relevance. This ordering of the effects reflects a reasonable ordering of the sorts of questions they answer. While following this line of thinking, we stumbled upon the broad class of interventional effects. This class includes, in addition to the interventional (in)direct effects, several other effect variations that are more intuitive and fit certain research questions better than the interventional (in)direct effects.

As the focus is on effect definition, we mostly put aside questions about whether an effect is identified and how it may be estimated, except a couple of identification comments called for by the juxtaposition of interventional and natural (in)direct effect types. The closing remarks provide brief comments that aim to orient the reader to these two topics, and refer the reader to the relevant literature.

This paper does not assume that the reader is well versed in causal inference reasoning. We use as plain as possible language combined with examples to elucidate concepts and ideas. While some mathematical notation is needed, it is accompanied by explanations in English and is color-coded for easy recognition. Also, the paper includes extensive footnotes. We recommend that the reader skip the footnotes on the first read of the paper, and come back to them on a second read. The footnotes are not required for understanding the main content, but provide additional explanations to deepen understanding, and expose a broader range of terms and concepts encountered in the causal mediation literature, which may help the reader be an effective consumer of that literature.

For simple presentation, we use the situation with a binary exposure, one mediator and one outcome. The content here applies directly to a non-binary exposure (replacing the exposed and unexposed conditions with the contrast of $A=a$ and $A=a^*$ where $a$ is the exposure level of interest and $a^*$ is the comparison level), multiple outcomes (treated as one outcome vector), and multiple mediators considered enbloc (as one mediator vector). Solid understanding of this case will make it easier to deal with more complex situations.

\section{Total effect}

\noindent Again, consider our intervention for high school students (the exposure) and their college readiness (the outcome). Each student (indexed by $i$) has two \textit{potential outcomes}, one is the college readiness level that the student would have if exposed to the intervention, the other is the college readiness level they would have if not exposed to the intervention. These are two different variables, labeled $\textcolor{red}{Y_i(1)}$ and $\textcolor{blue}{Y_i(0)}$, where 1 and 0 stand for the exposed and unexposed conditions. The \textit{individual total effect}, denoted ${T\!E}_i$, is often defined on the additive scale as the difference between these two potential outcomes, 
$${T\!E}_i:=\textcolor{red}{Y_i(1)}-\textcolor{blue}{Y_i(0)},$$
where the symbol $:=$ means ``is defined as''. Averaging the individual total effects over the population%
\footnote{For simplicity of presentation, we presume interest in average effects for the population. If the average effect of interest is that for the exposed, the unexposed, or another specific subpopulation, these expectations of the potential outcomes are taken conditional on the subpopulation.} 
of high school students, we have the \textit{average total effect}, denoted TE,
$$\mathrm{TE}:=\E[\textcolor{red}{Y(1)}]-\E[\textcolor{blue}{Y(0)}],$$
where $\E[\cdot]$ denotes expectation (population mean).%
\footnote{In this paper we focus on effects defined on the additive scale, the most common choice. Alternatively, causal effects may be defined on multiplicative scales, e.g., as mean ratios, rate ratios, risk ratios. A broader view is that the causal effect is the difference between the two potential outcomes' distributions.}
If the outcome is binary (coded 0/1), this definition is equivalent to $\mathrm{TE}=\P(\textcolor{red}{Y(1)}\!=\!1)-\P(\textcolor{blue}{Y(0)}\!=\!1)$, a risk difference.

\begin{table*}[ht!]
\caption{Individual total effects: a toy example}\label{tab:TE}
\centering
\resizebox{.65\textwidth}{!}{%
\begin{tabular}{cccccccccc}
\\
&& \multicolumn{2}{c}{Potential outcomes} && Total effect && \multicolumn{2}{c}{OBSERVED DATA} 
\\ \cline{3-4} \cline{6-6} \cline{8-9}
$i$ && \textcolor{blue}{$Y_i(0)$} & \textcolor{red}{$Y_i(1)$} && ${T\!E}_i=\textcolor{red}{Y_i(1)}-\textcolor{blue}{Y_i(0)}$ && $A_i$ & $Y_i$ 
\\ \hline
Bo && \textcolor{blue}{4} & \textcolor{red}{9} && 5 && 1 & \textcolor{red}{9} 
\\
Sam && \textcolor{blue}{7} & \textcolor{red}{8} && 1 && 1 & \textcolor{red}{8}
\\
Ian && \textcolor{blue}{5} & \textcolor{red}{7} && 2 && 1 & \textcolor{red}{7} 
\\
Ben && \textcolor{blue}{8} & \textcolor{red}{7} && -1 && 1 & \textcolor{red}{7} 
\\ \hline
Suri && \textcolor{blue}{3} & \textcolor{red}{5} && 2 && 0 & \textcolor{blue}{3} 
\\
Bill && \textcolor{blue}{6} & \textcolor{red}{7} && 1 && 0 & \textcolor{blue}{6}
\\
Kat && \textcolor{blue}{9} &  \textcolor{red}{8} && -1 && 0 & \textcolor{blue}{9} 
\\
Dre && \textcolor{blue}{4} & \textcolor{red}{8} && 4 && 0 & \textcolor{blue}{4}
\\ \hline
\\
\end{tabular}%
}
\end{table*}

To make things concrete, Table \ref{tab:TE} shows a toy example where the population consists of eight individuals. For each individual, both potential outcomes (values on a 0-10 scale) and the total effect are shown. TE is an average increase in college readiness of 1.625 points (the average of all the $TE_i$ values in the table). In reality, however, we are not privy to individual effects, for we never observe both potential outcomes for the same individual. What we observe is the realized outcome $Y_i$, which reveals one of the potential outcomes. Specifically, we observe $Y_i=\textcolor{red}{Y_i(1)}$ if the student is exposed to the intervention ($A_1=1$), and $Y_i=\textcolor{blue}{Y_i(0)}$ if the student is not exposed to the intervention ($A_i=0$).%
\footnote{This connection between the observed outcome and the potential outcomes is called the \textit{consistency} assumption \citep{VanderWeele2009}.}
The puzzle for researchers is that given the data in the last two columns of the table, we want to learn about the average total effect, and more.

\medskip

\begin{figure*}[h!]
\caption{The total effect}\label{fig:TE}
    \centering
\resizebox{.65\textwidth}{!}{%
\begin{tikzpicture}[
obs/.style={rectangle, minimum size=5mm},
unobs/.style={rectangle, minimum size=4mm},
setred/.style={rectangle, draw=red, thick, minimum size=5mm},
setblue/.style={rectangle, draw=blue, thick, minimum size=5mm}
]

\node[obs]   (C)                            {$C$};
\node[obs]   (A)  [right=of C, xshift=+3mm] {$A$};
\node[obs]   (Y)  [right=of A, xshift=+3mm] {$Y$};
\node[unobs] (UA) [below=of A, yshift=+7mm] {\footnotesize $U_A$};
\node[unobs] (UY) [below=of Y, yshift=+7mm] {\footnotesize $U_Y$};
\draw[->] (C) -- (A);
\draw[->] (C) .. controls +(up:15mm) and +(up:15mm)  .. (Y);
\draw[thick, ->] (A) -- (Y);
\draw[->]        (UA) -- (A);
\draw[->]        (UY) -- (Y);

\node[obs]     (C0)  [right=of Y, xshift=+20mm] {$C$};
\node[setblue] (A0)  [right=of C0]               {$A=\textcolor{blue}{0}$};
\node[obs]     (Y0)  [right=of A0]               {\textcolor{blue}{$Y(0)$}};
\node[unobs]   (UY0) [below=of Y0, yshift=+7mm]  {\footnotesize $U_Y$};
\draw[->] (C0) .. controls +(up:15mm) and +(up:15mm)  .. (Y0);
\draw[thick, ->] (A0) -- (Y0);
\draw[->]        (UY0) -- (Y0);

\node[obs]    (C1)  [below=of C0, yshift=-12mm] {$C$};
\node[setred] (A1)  [right=of C1]              {$A=\textcolor{red}{1}$};
\node[obs]    (Y1)  [right=of A1]              {\textcolor{red}{$Y(1)$}};
\node[unobs]  (UY1) [below=of Y1, yshift=+7mm] {\footnotesize $U_Y$};
\draw[->] (C1) .. controls +(up:15mm) and +(up:15mm)  .. (Y1);
\draw[thick, ->] (A1) -- (Y1);
\draw[->]        (UY1) -- (Y1);

\node[obs] [above right=of C, yshift=+4mm, xshift=-21mm]  {\begin{tabular}{c}a. The regular world \\ without any manipulation\end{tabular}};
\node[obs] [above right=of C0, yshift=+4mm, xshift=-25mm] {\begin{tabular}{c}b. The two worlds contrasted \\ by the total effect\end{tabular}};
\end{tikzpicture}}
\end{figure*}
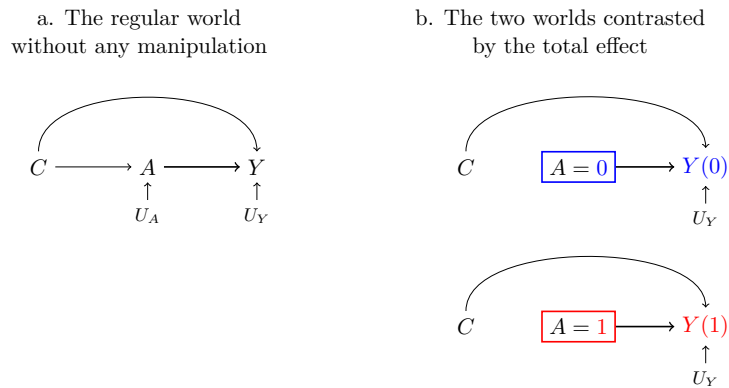

\noindent\textbf{Introducing the causal DAG.}
A causal DAG \citep{Pearl2009} consists of nodes that represent variables and arrows that represent causal effects; all common causes of any pair of variables are included; and there is no directed path from a variable (through other variables) to itself. Consider the DAG in Fig. \ref{fig:TE}a which shows the world without any (real or imagined) manipulation by the investigator. The variables of interest are $A$ (college prep program participation) and $Y$ (college readiness). $C$ represents \textit{common causes} of $A$ and $Y$ -- variables that may make a high school student more or less likely to attend a college preparation program and that influence college readiness (e.g., academic achievement, socio-economic status, etc.). Another name for a common cause is \textit{confounder}, so $C$ consists of confounders of the $A$-$Y$ relationship. This DAG also shows that $A$ and $Y$ have causes that are not shared, $U_A$ and $U_Y$ (where $U$ stands for \textit{unique causes}); these may also be left out of the DAG and their presence is implicitly understood.%
\footnote{Readers familiar with structural equation modeling must have noticed that this DAG looks identical to a structural equation model (SEM). In fact, a SEM represented by an identical diagram encodes the same assumptions as the DAG about which variables are or are not causes (or effects) of which variables. Typically, a SEM also assumes certain model forms (often linear) for the causal relationships and certain distributions (often normal) for the disturbance terms. A DAG does not encode such assumptions; it is nonparametric.}

The total effect is a contrast between two parallel worlds which we imagine for the same individual (or the same group of individuals). In these two worlds, everything is the same, except that in one world $A$ is set to 1 (program participation, i.e., the exposed condition), and in the other world $A$ is set to 0 (nonparticipation, i.e., the unexposed condition). These are shown in the two DAGs in Fig. \ref{fig:TE}b, where the box represents that a variable is set to a value. In these two DAGs, since $A$ is set to a fixed value for everyone, this breaks the causal link between $C$ and $A$.

\section{Natural (in)direct effects}

\noindent As mentioned earlier, there is more than one (in)direct effect type. To start, consider a common motivation of mediation analysis, the desire to explain the effect of the exposure on the outcome. The  question often asked is: How much (if any) of this effect went through the mediator, and how much went through other ways? \textit{Natural (in)direct effects} (\citeauthor{Robins1992}, \citeyear{Robins1992}; \citeauthor{Pearl2001}, \citeyear{Pearl2001})\footnote{\citeauthor{Robins1992} referred to these effects as \textit{pure} and \textit{total} (\textit{in})\textit{direct effects} (see more in foot note 9). \citeauthor{Pearl2001} labeled them \textit{natural} (\textit{in})\textit{direct effects}. \citeauthor{Pearl2001}'s label has become popular, and is useful in contrasting these effects with the \textit{interventional} (\textit{in})\textit{direct effects} to be presented in the next section.} capture the essence of this question.

The definition of these effects starts at the individual level. The idea is to split the individual total effect ${T\!E}_i$ -- the contrast of \textcolor{red}{$Y_i(1)$} and \textcolor{blue}{$Y_i(0)$} -- into two contrasts using a third potential outcome that is in-between in some sense, so that one contrast represents a mediated effect, the other a direct effect, for the individual. What should be this third potential outcome? The outcome in a \textit{hypothetical} world where the exposure is set to one condition, but the mediator is set to its potential value under the other exposure condition. (This world is a product of our imagination that cannot realize or be observed, but is conceptualized in order to help decompose the total effect into meaningful and somewhat interpretable pieces.)

To see how this works, we need to be a bit more formal. Let $Y_i(a)$ denote the potential outcome of individual $i$ if exposure is set to $a$ (where $a$ can be either 1 or 0, so any statement about $Y_i(a)$ applies to both $\textcolor{red}{Y_i(1)}$ and $\textcolor{blue}{Y_i(0)}$). Similarly, let $M_i(a)$ denote the potential mediator value for individual $i$ if exposure is set to $a$. Also, consider a new type of potential outcome, $Y_i(a,m)$, the potential outcome when exposure is set to $a$ and mediator is set to $m$. Depending on the mediator distribution, there may be many such potential outcomes -- two for each possible $m$ value. Yet we are not interested in just any $m$ value. For each individual, only two values are relevant: $\textcolor{red}{M_i(1)}$ and $\textcolor{blue}{M_i(0)}$, the values that the mediator would \textit{naturally} take under the two exposure conditions. This is because the mediated part of ${T\!E}_i$ occurs only because the exposure has an effect on the mediator, and that effect is the difference between $\textcolor{red}{M_i(1)}$ and $\textcolor{blue}{M_i(0)}$. Hence we replace $m$ in $Y_i(a,m)$ with one of these values, denoted $M_i(a')$ (where $a'$ can also be either 0 or 1), and obtain $Y_i(a,M_i(a'))$, the potential outcome in a world where the exposure is set to $a$ and the mediator is set to the value it would take under exposure $a'$.

When $a$ and $a'$ are the same, we have $Y_i(a,M_i(a))$, which is equal to $Y_i(a)$. That is, the potential outcome in a world where exposure is set to $a$ and mediator is set to $M_i(a)$ is the same as the potential outcome in a world where exposure is set to $a$ and mediator follows naturally.%
\footnote{This is called the \textit{composition} assumption \citep{VanderWeele2009}.} ${T\!E}_i$ is thus a shift from $\textcolor{blue}{Y_i(0)}=\textcolor{blue}{Y_i(0,M_i(0))}$ to $\textcolor{red}{Y_i(1)}=\textcolor{red}{Y_i(1,M_i(1))}$.

Our in-between world is one where $a$ and $a'$ are not the same. One choice is the world where the exposure is set to \textcolor{red}{1}, but the mediator is set, for each individual, to their $\textcolor{blue}{M_i(0)}$. With this as the in-between world, ${T\!E}_i$ is split into two parts: Part 1 is a shift from $\textcolor{blue}{Y_i(0,M_i(0))}$ to the in-between potential outcome $Y_i(\textcolor{red}{1},\textcolor{blue}{M_i(0)})$. This fits the notion of a direct effect (and is called a \textit{natural direct effect} (NDE)), as it is the effect of changing the exposure from $\textcolor{blue}{0}$ to $\textcolor{red}{1}$ but fixing the mediator (not letting the mediator change in response to the exposure change). Part 2 is a shift from $Y_i(\textcolor{red}{1},\textcolor{blue}{M_i(0)})$ to $\textcolor{red}{Y_i(1,M_i(1))}$. This fits the notion of an indirect effect (and is called a \textit{natural indirect effect} (NIE)), as it is the effect of the mediator switching from $\textcolor{blue}{M_i(0)}$ to $\textcolor{red}{M_i(1)}$ (as if in response to a change in exposure), while the exposure is actually fixed (so there is no direct effect element).

\begin{table*}[t]
\caption{Individual natural (in)direct effects of the direct-indirect decomposition: a toy example}\label{tab:NDENIE}
\centering
\resizebox{.8\textwidth}{!}{%
\begin{tabular}{cccccccccccccccccc}
\\
&& \multicolumn{2}{c}{Potential} && \multicolumn{3}{c}{Potential} && \multicolumn{2}{c}{Direct-indirect} && \multicolumn{3}{c}{OBSERVED} 
\\
&& \multicolumn{2}{c}{mediators} && \multicolumn{3}{c}{outcomes} && \multicolumn{2}{c}{decomposition} && \multicolumn{3}{c}{DATA}
\\ \cline{3-4} \cline{6-8}  \cline{10-11} \cline{13-15}
$i$ && \textcolor{blue}{$M_i(0)$} & \textcolor{red}{$M_i(1)$} && \textcolor{blue}{$Y_i(0)$} & $Y_i(\textcolor{red}{1},\textcolor{blue}{M_i(0)})$ & \textcolor{red}{$Y_i(1)$} && ${NDE}_i(\boldsymbol\cdot\textcolor{blue}0)$ & ${NIE}_i(\textcolor{red}{1}\boldsymbol\cdot)$ && $A_i$ & $M_i$ & $Y_i$
\\ \hline
Bo && \textcolor{blue}{low} & \textcolor{red}{high} && \textcolor{blue}{4} & 7 & \textcolor{red}{9} && 3 & 2 && 1 & \textcolor{red}{high} & \textcolor{red}{9}
\\
Sam && \textcolor{blue}{medium} & \textcolor{red}{high} && \textcolor{blue}{7} & 7 & \textcolor{red}{8} && 0 & 1 && 1 & \textcolor{red}{high} & \textcolor{red}{8}
\\
Ian && \textcolor{blue}{low} & \textcolor{red}{low} && \textcolor{blue}{5} & 7 & \textcolor{red}{7} && 2 & 0 && 1 & \textcolor{red}{low} & \textcolor{red}{7}
\\
Ben && \textcolor{blue}{low} & \textcolor{red}{medium} && \textcolor{blue}{8} & 6 & \textcolor{red}{7} && -2 & 1 && 1 & \textcolor{red}{medium} & \textcolor{red}{7}
\\ \hline
Suri && \textcolor{blue}{low} & \textcolor{red}{high} && \textcolor{blue}{3} & 3 & \textcolor{red}{5} && 0 & 2 && 0 & \textcolor{blue}{low} & \textcolor{blue}{3}
\\
Bill && \textcolor{blue}{low} & \textcolor{red}{medium} && \textcolor{blue}{6} & 5 & \textcolor{red}{7} && -1 & 2 && 0 & \textcolor{blue}{low} & \textcolor{blue}{6}
\\
Kat && \textcolor{blue}{high} & \textcolor{red}{high} && \textcolor{blue}{9} & 8 & \textcolor{red}{8} && 1 & 0 && 0 & \textcolor{blue}{high} & \textcolor{blue}{9}
\\
Dre && \textcolor{blue}{medium} & \textcolor{red}{high} && \textcolor{blue}{4} & 7 & \textcolor{red}{8} && 3 & 1 && 0 & \textcolor{blue}{medium} & \textcolor{blue}{4}
\\ \hline
\\
\end{tabular}%
}
\end{table*}

For concreteness, consider Bo, one of our high school students. Suppose we are omniscient and know what would happen in all the different worlds. If Bo participated in the college prep program, Bo would have high self-awareness and college readiness level 9 -- these are $\textcolor{red}{M_\text{Bo}(1)}$ and $\textcolor{red}{Y_\text{Bo}(1)}$. If Bo did not participate in the program, Bo would have low self-awareness and college readiness level 4 -- these are $\textcolor{blue}{M_\text{Bo}(0)}$ and $\textcolor{blue}{Y_\text{Bo}(0)}$. The total effect of the intervention for Bo is thus an increase in college readiness of 5 points. In the in-between world where Bo participated in the program but somehow Bo's self-awareness was fixed at its level under nonparticipation, $\textcolor{blue}{M_\text{Bo}(0)}$ (i.e., low), Bo would attain college readiness level 7 -- this is $Y_\text{Bo}(\textcolor{red}{1},\textcolor{blue}{M_\text{Bo}(0)})$. This means the NDE for Bo is an increase in college readiness of 3 points, from level 4 to 7, and the NIE is a further increase of 2 points, from level 7 to 9. Table \ref{tab:NDENIE} shows these effects for all the high-schoolers in our toy example.

We refer to the above decomposition of the total effect the \textit{direct-indirect} decomposition, based on the order of the component effects.  Alternatively, we can use the other in-between world with $Y_i(\textcolor{blue}{0},\textcolor{red}{M_i(1)})$ to split the total effect into a NIE followed by a NDE -- the \textit{indirect-direct} decomposition.

\medskip

The \textit{average} natural (in)direct effects, which are population means of the individual effects, decompose the \textit{average} total effect:
\begin{align*}
    &\text{direct-indirect decomposition:}
    \\
    &
    {\small\mathrm{TE}=\small\underbrace{\E[\textcolor{red}{Y(1,M(1))}]\!-\!\E[Y(\textcolor{red}{1},\textcolor{blue}{M(0)})]}_{\textstyle\mathrm{NIE}(\textcolor{red}{1}\boldsymbol{\cdot})}+\underbrace{\E[Y(\textcolor{red}{1},\textcolor{blue}{M(0)})]\!-\!\E[\textcolor{blue}{Y(0,M(0))}]}_{\textstyle\mathrm{NDE}(\boldsymbol{\cdot}\textcolor{blue}{0})},}
    \\
    &\text{indirect-direct decomposition:}
    \\
    &{\small\mathrm{TE}=\underbrace{\E[\textcolor{red}{Y(1,M(1))}]\!-\!\E[Y(\textcolor{blue}{0},\textcolor{red}{M(1)})]}_{\textstyle\mathrm{NDE}(\boldsymbol{\cdot}\textcolor{red}{1})}+\underbrace{\E[Y(\textcolor{blue}{0},\textcolor{red}{M(1)})]\!-\!\E[\textcolor{blue}{Y(0,M(0))}]}_{\textstyle\mathrm{NIE}(\textcolor{blue}{0}\boldsymbol{\cdot})}.}
\end{align*}
To differentiate between the two NDEs and the two NIEs,%
\footnote{Another way to differentiate these effects is the labeling of the first effect in a decomposition as ``pure'' and the second as ``total'' \citep{Robins1992}. In this labeling scheme, $\mathrm{NDE}(\boldsymbol{\cdot}\textcolor{blue}{0})$ and $\mathrm{NIE}(\textcolor{red}{1}\boldsymbol{\cdot})$, for example, are called the \textit{pure direct effect} and \textit{total indirect effect}. We opt not to use this terminology, as it is confusing to have ``total'' in both \textit{total (in)direct effect} and \textit{total effect}.}
here we index each of these effects with a combination of a dot representing the condition (either exposure or mediator) that varies in the contrast, and a number representing the condition that is fixed.%
\footnote{Here the dot is helpful but not necessary because the ``direct'' and ``indirect'' labels already imply which condition is varying. It is more important when considering the case with multiple mediators.}

\medskip

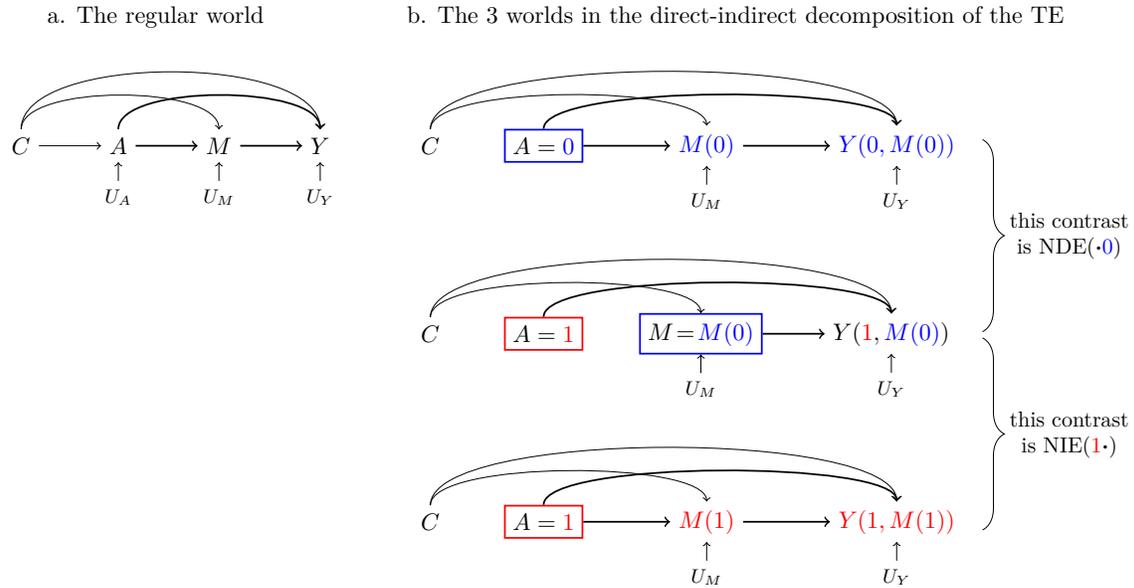
\begin{figure*}[t]
\caption{Natural (in)direct effects -- depicted in the special case with no intermediate confounder}\label{fig:NIDE}
    \resizebox{\textwidth}{!}{%
    \begin{tikzpicture}[
        obs/.style={rectangle, minimum size=5mm},
        unobs/.style={rectangle, minimum size=4mm},
        setred/.style={rectangle, draw=red, thick, minimum size=5mm},
        setblue/.style={rectangle, draw=blue, thick, minimum size=5mm}
    ]
        \node[obs]    (C)                            {$C$};
        \node[obs]    (A)  [right=of C] {$A$};
        \node[obs]    (M)  [right=of A] {$M$};
        \node[obs]    (Y)  [right=of M] {$Y$};
        \node[unobs]  (UA) [below=of A, yshift=+7mm] {\footnotesize $U_A$};
        \node[unobs]  (UM) [below=of M, yshift=+7mm] {\footnotesize $U_M$};
        \node[unobs]  (UY) [below=of Y, yshift=+7mm] {\footnotesize $U_Y$};
        \draw[->] (C) -- (A);
        \draw[->] (C) .. controls +(up:15mm) and +(up:15mm)  .. (Y);
        \draw[->] (C) .. controls +(up:10mm) and +(up:10mm)  .. (M);
        \draw[thick, ->] (A) -- (M);
        \draw[thick, ->] (A) .. controls +(up:10mm) and +(up:10mm)  .. (Y);
        \draw[thick, ->] (M) -- (Y);
        \draw[->]        (UA) -- (A);
        \draw[->]        (UM) -- (M);
        \draw[->]        (UY) -- (Y);

        \node[obs]     (C0)  [right=of Y, xshift=+2mm]   {$C$};
        \node[setblue]  (A0)  [right=of C0, xshift=-1mm] {$A=\textcolor{blue}{0}$};
        \node[obs] (M0)  [right=of A0, xshift=+4mm] {$\textcolor{blue}{M(0)}$};
        \node[obs]     (Y0)  [right=of M0, xshift=+4mm]              {$\textcolor{blue}{Y(0,M(0))}$};
        \node[unobs]   (UM0) [below=of M0, yshift=+7mm] {\footnotesize $U_M$};
        \node[unobs]   (UY0) [below=of Y0, yshift=+7mm] {\footnotesize $U_Y$};
        \draw[->] (C0) .. controls +(up:15mm) and +(up:15mm)  .. (Y0);
        \draw[->]        (C0) .. controls +(up:10mm) and +(up:10mm)  .. (M0);
        \draw[thick, ->] (A0) -- (M0);
        \draw[thick, ->] (A0) .. controls +(up:10mm) and +(up:10mm)  .. (Y0);
        \draw[thick, ->] (M0) -- (Y0);
        \draw[->]        (UM0) -- (M0);
        \draw[->]        (UY0) -- (Y0);

        \node[obs]     (C10)  [below=of C0, yshift=-15mm]   {$C$};
        \node[setred]  (A10)  [right=of C10, xshift=-1mm] {$A=\textcolor{red}{1}$};
        \node[setblue] (M10)  [right=of A10, xshift=-1mm] {$M\!=\!\textcolor{blue}{M(0)}$};
        \node[obs]     (Y10)  [right=of M10] {$Y(\textcolor{red}{1},\textcolor{blue}{M(0)})$};
        \node[unobs]   (UM10) [below=of M10, yshift=+7mm] {\footnotesize $U_M$};
        \node[unobs]   (UY10) [below=of Y10, yshift=+7mm] {\footnotesize $U_Y$};
        \draw[->] (C10) .. controls +(up:15mm) and +(up:15mm)  .. (Y10);
        \draw[->]        (C10) .. controls +(up:10mm) and +(up:10mm)  .. (M10);
        \draw[thick, ->] (A10) .. controls +(up:10mm) and +(up:10mm)  .. (Y10);
        \draw[thick, ->] (M10) -- (Y10);
        \draw[->]        (UM10) -- (M10);
        \draw[->]        (UY10) -- (Y10);

        \node[obs]     (C1)  [below=of C10, yshift=-15mm]   {$C$};
        \node[setred]  (A1)  [right=of C1, xshift=-1mm] {$A=\textcolor{red}{1}$};
        \node[obs] (M1)  [right=of A1, xshift=+4mm] {$\textcolor{red}{M(1)}$};
        \node[obs]     (Y1)  [right=of M1, xshift=+4mm]              {$\textcolor{red}{Y(1,M(1))}$};
        \node[unobs]   (UM1) [below=of M1, yshift=+7mm] {\footnotesize $U_M$};
        \node[unobs]   (UY1) [below=of Y1, yshift=+7mm] {\footnotesize $U_Y$};
        \draw[->] (C1) .. controls +(up:15mm) and +(up:15mm)  .. (Y1);
        \draw[->]        (C1) .. controls +(up:10mm) and +(up:10mm)  .. (M1);
        \draw[thick, ->] (A1) -- (M1);
        \draw[thick, ->] (A1) .. controls +(up:10mm) and +(up:10mm)  .. (Y1);
        \draw[thick, ->] (M1) -- (Y1);
        \draw[->]        (UM1) -- (M1);
        \draw[->]        (UY1) -- (Y1);

        \node[obs] [above right=of C, yshift=+5mm, xshift=-10mm] {a. The regular world};
        \node[obs] [above right=of C0, yshift=+5mm, xshift=-18mm] {b. The 3 worlds in the direct-indirect decomposition of the TE};
        \node[obs] [above=of C, yshift=10mm] {};

        \draw[decorate,decoration={brace,amplitude=10pt},xshift=0pt,yshift=0pt] (15.5,0.1) -- (15.5,-3.0) node [black,midway,xshift=1.4cm] {\small\begin{tabular}{c}this contrast\\is $\mathrm{NDE}(\boldsymbol\cdot\textcolor{blue}{0})$\end{tabular}};
        \draw[decorate,decoration={brace,amplitude=10pt},xshift=0pt,yshift=0pt] (15.5,-3.1) -- (15.5,-6.2) node [black,midway,xshift=1.4cm] {\small\begin{tabular}{c}this contrast\\is $\mathrm{NIE}(\textcolor{red}{1}\boldsymbol\cdot)$\end{tabular}};
    \end{tikzpicture}}
\end{figure*}

\noindent\textbf{Revisiting the causal DAG.}
Fig. \ref{fig:NIDE} shows the direct-indirect decomposition of the total effect. Fig. \ref{fig:NIDE}a is the unmanipulated DAG. In Fig. \ref{fig:NIDE}b, the top and bottom DAGs represent the two worlds contrasted by the total effect, where the mediator follows naturally after the exposure. The middle DAG in Fig. \ref{fig:NIDE}b represents the in-between world, where the mediator is set to its value under the opposite exposure condition.%
\footnote{While there is no arrow from $C$ to \fcolorbox{red}{white}{$A=\textcolor{red}{1}$} as $A$ is fixed to a constant for every individual, there is an arrow from $C$ to \fcolorbox{blue}{white}{$M=\textcolor{blue}{M(0)}$} because $\textcolor{blue}{M(0)}$ values vary across individuals and depend of $C$ values.}

Note that $C$ in Fig. \ref{fig:NIDE} is different from $C$ in Fig. \ref{fig:TE}. Here $C$ collectively represents three sets of confounders, of the $A$-$M$, $A$-$Y$ and $M$-$Y$ relationships, which may overlap but may not be identical. Note also that this figure represents the special (and simple) case where no $M$-$Y$ confounders are influenced by $A$. In the general case (see Fig. \ref{fig:SIDE}a), in addition to $M$-$Y$ confounders not influenced by $A$ (included in $C$), there are $M$-$Y$ confounders influenced by $A$ (termed \textit{intermediate confounders}) represented by $L$.%
\footnote{In the presence of $L$, $Y(a,M(a'))$ is equivalent to $Y(a,L(a),M(a'))$, because $L$ follows naturally after exposure $a$ and thus takes on value $L(a)$, but $M$ is set to value $M(a')$.}

\medskip

In our experience, when first introduced, the natural (in)direct effects may take some time to sink in. We offer two heuristics that might help build intuition for these effects.

\smallskip

\noindent\textbf{Heuristic definition 1: information flows.} Consider a system with three variables, exposure, mediator and outcome; and two paths along which information can flow in the system: the exposure $\to$ outcome path, and the exposure $\to$ mediator $\to$ outcome path. In the default condition of the system, called the \textit{unexposed condition}, 
$$\text{exposure}=\textcolor{blue}{0},~~ \text{mediator}=\textcolor{blue}{M(0)},~~ \text{outcome}=\textcolor{blue}{Y(0,M(0))},$$
and there is no information movement. If we switch the exposure from \textcolor{blue}{0} to \textcolor{red}{1}, this change is information. This information flows through both paths, and affects any variable it comes across. This results in a switch of both the mediator and the outcome, obtaining a new condition for the system, called the \textit{exposed condition}, where
$$\text{exposure}=\textcolor{red}{1},~~ \text{mediator}=\textcolor{red}{M(1)},~~ \text{outcome}=\textcolor{red}{Y(1,M(1))}.$$
The outcome switch from the unexposed to the exposed condition is the total effect.

Decomposition of the total effect is analogous to staggering the flow of information -- by turning a path off and on. Suppose we turn off the exposure $\to$ mediator $\to$ outcome path. When the exposure is switched from \textcolor{blue}{0} to \textcolor{red}{1}, the generated information can flow only through the exposure $\to$ outcome path. It does not reach, therefore does not affect, the mediator. But it reaches the outcome and causes it to change. As a result, the system obtains a \textit{mid-way condition}, where 
$$\text{exposure}=\textcolor{red}{1},~~ \text{mediator}=\textcolor{blue}{M(0)},~~ \text{outcome}=Y(\textcolor{red}{1},\textcolor{blue}{M(0)}).$$
Now we turn the exposure $\to$ mediator $\to$ outcome path back on. Information now flows through this path. It reaches the mediator, causing it to change. This flow of information then reaches the outcome, causing the outcome to change. This takes the system to the exposed condition. The two outcome shifts, from the unexposed to the mid-way condition, and from the mid-way to the exposed condition, are the $\mathrm{NDE}(\boldsymbol\cdot\textcolor{blue}{0})$ and $\mathrm{NIE}(\textcolor{red}{1}\boldsymbol\cdot)$.

If instead of turning off and on the exposure $\to$ mediator $\to$ outcome path, we turn off and on the exposure $\to$ outcome path, then we have an alternative mid-way condition where $\text{mediator}=\textcolor{red}{M(1)}$ and $\text{outcome}=Y(\textcolor{blue}{0},\textcolor{red}{M(1)})$. In this case, the two outcome shifts correspond to the $\mathrm{NIE}(\textcolor{blue}{0}\boldsymbol\cdot)$ and $\mathrm{NDE}(\boldsymbol\cdot\textcolor{red}{1})$.

\medskip

\noindent\textbf{Heuristic definition 2: double exposure.} Another way to think about natural (in)direct effects is to imagine that the exposure $A$ is the combination of two exposures denoted $A^M$ and $A^R$. A switch of $A^M$ from 0 to 1 causes a switch of the mediator from $M(0)$ to $M(1)$, so $A^M$ is responsible for any influence of $A$ on the outcome through the mediator. $A^R$ is responsible for all the remaining influence of $A$. We only observe situations where these two exposures go together, $A^R=A^M=A=1$ or $=0$. But imagine that they do not need to go together. With two exposures, we have potential outcomes of the form $Y(a,a')$, the outcome that would occur if $A^R$ were set to $a$ and $A^M$ to $a'$. The total effect previously defined corresponds to the effect of both exposures combined,
$$\mathrm{TE}=\E[Y(\textcolor{red}{1},\textcolor{red}{1})]-\E[Y(\textcolor{blue}{0},\textcolor{blue}{0})].$$
Its direct-indirect decomposition is given by adding and subtracting $\E[Y(\textcolor{red}{1},\textcolor{blue}{0})]$, i.e.,
$$\mathrm{TE}=\underbrace{\E[Y(\textcolor{red}{1},\textcolor{red}{1})]-\E[Y(\textcolor{red}{1},\textcolor{blue}{0})]}_{\textstyle\mathrm{NIE}(\textcolor{red}{1}\boldsymbol{\cdot})}\,+\,\underbrace{\E[Y(\textcolor{red}{1},\textcolor{blue}{0})]-\E[Y(\textcolor{blue}{0},\textcolor{blue}{0})]}_{\textstyle\mathrm{NDE}(\boldsymbol{\cdot}\textcolor{blue}{0})},$$
and its indirect-direct decomposition is given by adding and subtracting $\E[Y(\textcolor{blue}{0},\textcolor{red}{1})]$. In this representation, the two NDEs are the effects of $A^R$ when $A^M$ is set to 0 and to 1, and the two NIEs are the effects of $A^M$ when $A^R$ is set to 0 and to 1.

\medskip

\noindent\textbf{{Why two decompositions and which one to choose?}} As mentioned above, natural (in)direct effects match the purpose of explaining the causal effect, answering the question what part of the total effect went through the mediator and what part did not. These effects have thus been called ``descriptive'' \citep{Pearl2001}. A seemingly complicating matter is that with two decompositions of the total effect into natural (in)direct effects, there is not a single NDE or a single NIE. Generally, $\mathrm{NIE}(\textcolor{red}{1}\boldsymbol\cdot)$ and $\mathrm{NIE}(\textcolor{blue}{0}\boldsymbol\cdot)$ are not the same, and neither are $\mathrm{NDE}(\boldsymbol\cdot\textcolor{blue}{0})$ and $\mathrm{NDE}(\boldsymbol\cdot\textcolor{red}{1})$. This is not a weakness of these effects. On the contrary, it reveals an implicit assumption in our original question, that the total effect can be split into two effects that are separate and do not interact -- which, in the language of the double exposure heuristic definition, means the effect of $A^R$ must be constant for both levels of $A^M$ and vice versa. This is an arbitrary assumption. 

A practical question remains: For a specific analysis, which decomposition should be used? Or should both be used? Our view is that answering this question requires being more clear about the research question/objective. We propose three answers for three cases.

Case 1: \textit{Is there a mediated effect? Or, is the causal effect (partly) mediated by this putative mediator? (And if yes, what is the size of the effect?)} With this research question, we propose using the direct-indirect decomposition. The rationale is that here we are not questioning the existence of a direct effect, but are considering the possibility of a mediated effect in addition to the direct effect; if there is no mediated effect (either because the mediator does not change in response to the exposure, or it does change but that change does not cause a change in the outcome), then the total effect is the same as $\mathrm{NDE}(\boldsymbol\cdot\textcolor{blue}{0})$, the direct effect in the direct-indirect decomposition.

Case 2: \textit{In addition to the mediated effect, is there a direct effect? Or, does the exposure influence the outcome in other ways, not through this mediator? (And if yes, what is the size of this effect?)} This is a mirror image of the previous case, just flipping the relative order of direct and indirect effects. We propose using the indirect-direct decomposition. If there is no direct effect, then the total effect is the same as $\mathrm{NIE}(\textcolor{blue}{0}\boldsymbol\cdot)$, the indirect effect in the indirect-direct decomposition.

Case 3: \textit{The objective is to describe the effect of exposure on outcome with direct and indirect effect elements, without any prior assumption or preferred question about either direct or indirect effects.} In this case, we propose presenting both decompositions. After all, if the purpose is simply to describe all we can learn, there is no reason to prefer either decomposition over the other.

While all three cases may be encountered in research practice, we suspect that case 1 is more common than case 2. This may explain why methodological papers tend to either present both decompositions or present only the direct-indirect decomposition; we have not come across any that presents only the indirect-direct decomposition.

\section{Interventional effects}

\noindent The natural (in)direct effects are defined based on an \textit{explanatory} perspective -- motivated by a desire to explain the total effect. There are times when researchers may approach a mediation setting with an \textit{interventional} perspective -- asking what if we could modify the exposure, or intervene on the mediator, in a certain way.

A situation where one may ask questions of this type is where the goal of the research program is to develop, or to contribute to the development of, an effective intervention. This may take multiple rounds of development and modification. Consider our college prep program study as one step in the development process. In addition to testing the current form of the program, investigators might ask, what would be the effect of the program (i) if program components that only serve to improve self-awareness were eliminated, (ii) if only self-awareness-related components were kept, or (iii) if some other modification were made. If data from the current study shed some light on these questions, that informs decisions about whether to keep the program as is, or what modification should be made. A new version of the program, if created, is to be tested in a next study.

In other cases, one may ask the question \textit{what if we could intervene on an exposure or mediator} separately from questions about whether such an intervention is possible or what form it might take. This counterfactual thinking is relevant to research on health/social disparities, as it helps us imagine alternative worlds where social and structural elements that contribute to disparities were mitigated or neutralized \citep{Glymour2014,Jackson2018}; we will discuss one specific example of this in this section.

With this interventional, \textit{what if}, perspective, researchers should pay attention to a different class of effects, the \textit{interventional effects}. Perhaps the most well-known interventional effects in the mediation setting are \textit{interventional (in)direct effects} \citep{Didelez2006,VanderWeele2014a,Lok2016,Vansteelandt2017}.%
\footnote{It seems that these effects first appeared in \cite{Didelez2006}; this paper discussed these effects as a modification of the natural (in)direct effects, defined based on intervention regimes. The labeling of these effects as \textit{interventional} effects emerged out of a collection of several labels used in later papers, including ``effects of organic \textit{interventions}'' \citep{Lok2016} and ``randomized \textit{interventional} analogs of the natural (in)direct effects'' \citep{Vansteelandt2017}. The label \textit{interventional (in)direct effects} emphasizes their interpretation as effects of (hypothetical) interventions. Some authors also use the label \textit{stochastic effects}, short for \textit{stochastic interventional effects} \citep[e.g.,][]{Rudolph2017}, highlighting that mediator values are random in the conditions contrasted (this will be clear shortly).\label{fn:stochastic}}
There are also other \textit{interventional effect} variations that are relevant to some research questions.

\medskip

\subsection{Interventional (in)direct effects}

\noindent These effects inherit from the natural effects the same notions of \textit{direct} (not involving the mediator) and \textit{indirect} (involving a mediator shift that is related to a shift in exposure) effects. The label \textit{interventional}, however, means that the conditions contrasted correspond to \textit{interventions} on the treatment and/or mediator \citep{Didelez2006} that are conceivable -- for a hypothetical future study.%
\footnote{This feature is also referred to as \textit{experimental testability} \citep{Didelez2006} or \textit{manipulability} \citep{Robins2003}. Note that this concerns a principal possibility, not practical feasibility or ethical acceptability.}
Natural (in)direct effects do not belong in this class of effects because, absent time travel or magical knowledge of unobserved values, \textit{no intervention} can bring into existence either of the in-between worlds described earlier. Even if we could intervene on the mediator and make it take on any value we choose, after setting exposure to \textcolor{red}{1}, we cannot set the mediator to the individual-specific $\textcolor{blue}{M(0)}$ value because  we do not get to know what value it is.

We describe the \textit{interventional} (\textit{in})\textit{direct effects} from both the individual and the population point of view, as different readers may find one or the other more meaningful.

Let's start from the viewpoint of the individual, which highlights how these effects differ from their natural counterparts. Recall the potential outcomes that define the natural effects: $Y(a,M(a'))$ is the outcome in a world where the exposure is set to $a$ and the mediator is set, for the individual, to their own potential mediator value under exposure $a'$. Now, imagine that the individual is not assigned this specific value. Instead, the individual is grouped with others in a subpopulation with the same value/pattern on covariates $C$, i.e., the confounders not influenced by $A$ (we will comment on this shortly). Within this subpopulation, all the $M(a')$ values are put in a pool, and the individual is assigned a value randomly drawn from this pool.%
\footnote{The label \textit{stochastic effects} mentioned in footnote \ref{fn:stochastic} reflects this idea that mediator values are randomly drawn, as \textit{stochastic} means random (based on a distribution).}
Such pools (one for each $C$ value) constitute what is called \textit{the distribution of $M(a')$ conditional on $C$}, which we abbreviate to $d_{M(a')|C}$ ($d$ stands for \textit{distribution}, and the symbol $|C$ reads \textit{conditional on $C$} or \textit{given $C$}). The random draw from the subpopulation pool of $M(a')$ values is a draw from this distribution; we denote the draw by $\mathcal{M}(a'|C)$, with a squiggly letter $\mathcal{M}$ for randomness.

The interventional (in)direct effects are obtained by replacing $M(a')$ in the average (not individual) natural effect formulas with $\mathcal{M}(a'|C)$ \citep{Didelez2006, VanderWeele2014a}. There are two interventional direct effects
\begin{align*}
    \mathrm{IDE}(\boldsymbol{\cdot}\textcolor{blue}{0}):=\E[Y(\textcolor{red}{1},\mathcal{M}(\textcolor{blue}{0}|C))]-\E[Y(\textcolor{blue}{0},\mathcal{M}(\textcolor{blue}{0}|C))],
    \\
    \mathrm{IDE}(\boldsymbol{\cdot}\textcolor{red}{1}):=\E[Y(\textcolor{red}{1},\mathcal{M}(\textcolor{red}{1}|C))]-\E[Y(\textcolor{blue}{0},\mathcal{M}(\textcolor{red}{1}|C))],
\end{align*}
and two interventional indirect effects
\begin{align*}
    \mathrm{IIE}(\textcolor{blue}{0}\boldsymbol{\cdot}):=\E[Y(\textcolor{blue}{0},\mathcal{M}(\textcolor{red}{1}|C))]-\E[Y(\textcolor{blue}{0},\mathcal{M}(\textcolor{blue}{0}|C))],
    \\
    \mathrm{IIE}(\textcolor{red}{1}\boldsymbol{\cdot}):=\E[Y(\textcolor{red}{1},\mathcal{M}(\textcolor{red}{1}|C))]-\E[Y(\textcolor{red}{1},\mathcal{M}(\textcolor{blue}{0}|C))].
\end{align*}
Because the individual's $Y_i(a,\mathcal{M}(a'|C))$ depends on the random quantity $\mathcal{M}(a'|C)$, it is not meaningful to talk about interventional (in)direct effects for the individual. Unlike natural effects which are defined starting at the individual level, interventional (in)direct effects are essentially defined starting at the level of subpopulations defined by $C$ values.%
\footnote{Even though $C$ appears in the definition of interventional and not natural (in)direct effects, both effect types are conditional. Interventional (in)direct effects are conditional on $C$ (defined for each subpopulation with shared covariate values); natural effects are conditional on $i$ (defined for each individual).}

\begin{figure*}[t]
\caption{Interventional (in)direct effects -- depicted in the general case with an intermediate confounder}\label{fig:SIDE}
    \begin{center}
    \resizebox{\textwidth}{!}{%
    \begin{tikzpicture}[
        obs/.style={rectangle, minimum size=5mm},
        unobs/.style={rectangle, minimum size=4mm},
        setred/.style={rectangle, draw=red, thick, minimum size=5mm},
        setblue/.style={rectangle, draw=blue, thick, minimum size=5mm}
    ]
        \node[obs]  (C)    {$C$};
        \node[obs]  (A)  [right=of C, xshift=-2mm] {$A$};
        \node[obs]  (L)  [right=of A, xshift=-2mm] {$L$};
        \node[obs]  (M)  [right=of L, xshift=-2mm] {$M$};
        \node[obs]  (Y)  [right=of M, xshift=-2mm] {$Y$};
        \node[unobs]  (UA) [below=of A, yshift=+7mm] {\footnotesize $U_A$};
        \node[unobs]  (UL) [below=of L, yshift=+7mm] {\footnotesize $U_L$};
        \node[unobs]  (UM) [below=of M, yshift=+7mm] {\footnotesize $U_M$};
        \node[unobs]  (UY) [below=of Y, yshift=+7mm] {\footnotesize $U_Y$};
        \draw[->] (C) -- (A);
        \draw[->] (C) .. controls +(up:10mm) and +(up:10mm)  .. (L);
        \draw[->] (C) .. controls +(up:15mm) and +(up:15mm)  .. (M);
        \draw[->] (C) .. controls +(up:20mm) and +(up:20mm)  .. (Y);
        \draw[thick, ->] (A) -- (L);
        \draw[thick, ->] (A) .. controls +(up:10mm) and +(up:10mm)  .. (M);
        \draw[thick, ->] (A) .. controls +(up:15mm) and +(up:15mm)  .. (Y);
        \draw[thick, ->] (L) -- (M);
        \draw[thick, ->] (L) .. controls +(up:10mm) and +(up:10mm)  .. (Y);
        \draw[thick, ->] (M) -- (Y);
        \draw[->] (UA) -- (A);
        \draw[->] (UL) -- (L);
        \draw[->] (UM) -- (M);
        \draw[->] (UY) -- (Y);

        \node[obs]     (C00)  [right=of Y]   {$C$};
        \node[setblue] (A00)  [right=of C00, xshift=-4mm] {$A=\textcolor{blue}{0}$};
        \node[obs]     (L00)  [right=of A00, xshift=-2mm] {\textcolor{blue}{$L(0)$}};
        \node[setblue] (M00)  [right=of L00, xshift=-4mm] {$M\!=\!\mathcal{M}(\textcolor{blue}{0}|C)$};
        \node[obs]     (Y00)  [right=of M00, xshift=-3mm] {$Y(\textcolor{blue}{0},\mathcal{M}(\textcolor{blue}{0}|C))$};
        \node[unobs]   (UL00) [below=of L00, yshift=+7mm] {\footnotesize $U_L$};
        \node[unobs]   (UY00) [below=of Y00, yshift=+7mm] {\footnotesize $U_Y$};
        \draw[->] (C00) .. controls +(up:10mm) and +(up:10mm)  .. (L00);
        \draw[->]        (C00) .. controls +(up:15mm) and +(up:15mm)  .. (M00);
        \draw[->] (C00) .. controls +(up:20mm) and +(up:20mm)  .. (Y00);
        \draw[thick, ->] (A00) -- (L00);
        \draw[thick, ->] (A00) .. controls +(up:15mm) and +(up:15mm)  .. (Y00);
        \draw[thick, ->] (L00) .. controls +(up:10mm) and +(up:10mm)  .. (Y00);
        \draw[thick, ->] (M00) -- (Y00);
        \draw[->] (UL00) -- (L00);
        \draw[->] (UY00) -- (Y00);

        \node[obs]     (C10)  [below=of C00, yshift=-17mm]   {$C$};
        \node[setred]  (A10)  [right=of C10, xshift=-4mm] {$A=\textcolor{red}{1}$};
        \node[obs]     (L10)  [right=of A10, xshift=-2mm] {\textcolor{red}{$L(1)$}};
        \node[setblue] (M10)  [right=of L10, xshift=-4mm] {$M\!=\!\mathcal{M}(\textcolor{blue}{0}|C)$};
        \node[obs]     (Y10)  [right=of M10, xshift=-3mm] {$Y(\textcolor{red}{1},\mathcal{M}(\textcolor{blue}{0}|C))$};
        \node[unobs]   (UL10) [below=of L10, yshift=+7mm] {\footnotesize $U_L$};
        \node[unobs]   (UY10) [below=of Y10, yshift=+7mm] {\footnotesize $U_Y$};
        \draw[->] (C10) .. controls +(up:10mm) and +(up:10mm)  .. (L10);
        \draw[->]        (C10) .. controls +(up:15mm) and +(up:15mm)  .. (M10);
        \draw[->] (C10) .. controls +(up:20mm) and +(up:20mm)  .. (Y10);
        \draw[thick, ->] (A10) -- (L10);
        \draw[thick, ->] (A10) .. controls +(up:15mm) and +(up:15mm)  .. (Y10);
        \draw[thick, ->] (L10) .. controls +(up:10mm) and +(up:10mm)  .. (Y10);
        \draw[thick, ->] (M10) -- (Y10);
        \draw[->] (UL10) -- (L10);
        \draw[->] (UY10) -- (Y10);

        \node[obs]     (C11)  [below=of C10, yshift=-17mm]   {$C$};
        \node[setred]  (A11)  [right=of C11, xshift=-4mm] {$A=\textcolor{red}{1}$};
        \node[obs]     (L11)  [right=of A11, xshift=-2mm] {\textcolor{red}{$L(1)$}};
        \node[setred]  (M11)  [right=of L11, xshift=-4mm] {$M\!=\!\mathcal{M}(\textcolor{red}{1}|C)$};
        \node[obs]     (Y11)  [right=of M11, xshift=-3mm] {$Y(\textcolor{red}{1},\mathcal{M}(\textcolor{red}{1}|C))$};
        \node[unobs]   (UL11) [below=of L11, yshift=+7mm] {\footnotesize $U_L$};
        \node[unobs]   (UY11) [below=of Y11, yshift=+7mm] {\footnotesize $U_Y$};
        \draw[->] (C11) .. controls +(up:10mm) and +(up:10mm)  .. (L11);
        \draw[->]        (C11) .. controls +(up:15mm) and +(up:15mm)  .. (M11);
        \draw[->] (C11) .. controls +(up:20mm) and +(up:20mm)  .. (Y11);
        \draw[thick, ->] (A11) -- (L11);
        \draw[thick, ->] (A11) .. controls +(up:15mm) and +(up:15mm)  .. (Y11);
        \draw[thick, ->] (L11) .. controls +(up:10mm) and +(up:10mm)  .. (Y11);
        \draw[thick, ->] (M11) -- (Y11);
        \draw[->] (UL11) -- (L11);
        \draw[->] (UY11) -- (Y11);

        \node[obs] [above right=of C, yshift=+7mm, xshift=-7mm] {a. The regular world};
        \node[obs] [above right=of C00, yshift=+7mm, xshift=-10mm] {b. Intervention conditions defining an IDE and an IIE};

        \draw[decorate,decoration={brace,amplitude=10pt},xshift=0pt,yshift=0pt] (18,0.1) -- (18,-3.2) node [black,midway,xshift=1.5cm] {\small\begin{tabular}{l}this contrast\\gives $\mathrm{IDE}(\boldsymbol\cdot\textcolor{blue}{0})$\end{tabular}};
        \draw[decorate,decoration={brace,amplitude=10pt},xshift=0pt,yshift=0pt] (18,-3.3) -- (18,-6.5) node [black,midway,xshift=1.5cm] {\small\begin{tabular}{l}this contrast\\gives $\mathrm{IIE}(\textcolor{red}{1}\boldsymbol\cdot)$\end{tabular}};
    \end{tikzpicture}}
    \end{center}
    \caption*{\footnotesize Figure note: An equivalent representation of $Y(a,\mathcal{M}(a'|C))$ is $Y(a,L(a),\mathcal{M}(a'|C))$.}
\end{figure*}

Zooming all the way out to the viewpoint of the population, the effects just defined are contrasts of interventions that set the exposure and \textit{the mediator distribution}. (Intuitively, setting the mediator distribution to a specific distribution is the same as assigning each individual a random value drawn from that specific distribution.) Thus $\mathrm{IDE}(\boldsymbol\cdot a)$ is the effect of an exposure shift from \textcolor{blue}{0} to \textcolor{red}{1}, while the mediator distribution is set to $d_{M(a)|C}$, hence a direct effect. $\mathrm{IIE}(a\boldsymbol\cdot)$ is the effect of a mediator distribution shift from $d_{\textcolor{blue}{M(0)}|C}$ to $d_{\textcolor{red}{M(1)}|C}$ (as if in response to an exposure shift), while exposure is set to $a$, hence an indirect effect.%
\footnote{The equivalence of (i) assigning random mediator values to the individuals (seen from an individual perspective) and (ii) intervening on the mediator distribution (seen from a population perspective) is reflected in the common practice (in the literature) using the same notation ($G_{a|C}$, or $G_{M(a)|C}$, or $G_{M|a,C}$) to denote both the random draw and the distribution, which we here denote separately by $\mathcal{M}(a|C)$ and $d_{M(a)|C}$ for clarity of argument.}%
\textsuperscript{,}%
\footnote{One might ask whether effects can be defined based on different distribution for the mediator, for example, the marginal distribution (using $d_{M(a)}$ instead of $d_{M(a)|C}$). The answer is yes IF the research question is specifically about interventions that target such marginal distributions -- see section \textit{Interventional effects more generally} where we argue that effects should be defined to best match research questions. For a case (in health/social disparity research) where using marginal distributions is appropriate, see \cite{Jackson2019}. It should be noted that the standard definition that conditions on covariates has (i) a conceptual rationale (using the distribution that is relevant to the units, e.g., not assigning an elderly person a mediator value from a young person); and (ii) a technical convenience (one of the assumptions required for effect identification -- positivity -- is more likely to hold for $d_{M(a)|C}$ than for $d_{M(a)}$). To avoid ambiguity, we reserve the terms ``interventional (in)direct effects'' to refer to the (more or less standard) definition that conditions on $C$. Note that (again, depending on the research question) effects may also be defined where the distribution that the mediator is set to conditions on both pre-exposure covariates and on the intermediate confounder values that follow after exposure.}
In our current example, $\mathrm{IIE}(0\boldsymbol\cdot)$ might represent the effect on college readiness of a different intervention (possibly a modified version of the current program) if this intervention would shift the self-awareness distribution from $d_{\textcolor{blue}{M(0)}\mid C}$ to $d_{\textcolor{red}{M(1)}\mid C}$, and not change anything else that may affect college readiness.

We now briefly go over the relevant causal DAG and mention several properties of these effects before getting back to examples to make these effects more concrete. 

\medskip

\noindent\textbf{The causal DAG and the interventional mediator distribution.} Fig. \ref{fig:SIDE}b shows the intervention conditions that define $\mathrm{IDE}(\boldsymbol\cdot\textcolor{blue}{0})$ and $\mathrm{IIE}(\textcolor{red}{1}\boldsymbol\cdot)$. The distribution of $M$ (the mediator of interest) is set to $d_{\textcolor{blue}{M(0)}|C}$ in the top two conditions, and $d_{\textcolor{red}{M(1)}|C}$ in the bottom condition. The intermediate confounder $L$ (also a mediator), on the other hand, follows naturally after the exposure. This figure reflects the general case. In the special case with no intermediate confounders, the DAG simplifies, removing all $L$ related elements.

Note that among variables in $C$, not all may have a direct influence on $M$ ($A$-$M$, $L$-$M$ and $M$-$Y$ confounders do, but some $A$-$L$, $A$-$Y$ and $L$-$Y$ confounders may not). Therefore the distribution of $M(a)$ given $C$ is the same as the distribution of $M(a)$ given $C_{\!\mbox{\tiny\itshape M}}$ where $C_{\!\mbox{\tiny\itshape M}}$ is the subset of $C$ that has direct influence on $M$. Therefore the interventional (in)direct effects may be equivalently defined using $C_{\!\mbox{\tiny\itshape M}}$ \citep[as in][]{Didelez2006}.

\medskip

\noindent\textbf{Some properties that differ between interventional and natural (in)direct effects} should be noted.
First, unlike natural effects, interventional (in)direct effects do not decompose the total effect. This is not a limitation, as these effects were not created to explain the total effect.%
\footnote{This is a point where we disagree with some other authors who have commented that a drawback of these effects is that they do not sum to the total effect. Our opinion is, if the interest is in explaining TE (reflected in the desire for effects that sum to TE), then the logical target is the natural (in)direct effects. Interventional effects do not explain TE; they answer questions about (hypothetical) interventions.}
The four interventional (in)direct effects form two pairs that share the same sum, $\mathrm{IDE}(\boldsymbol\cdot\textcolor{blue}{0})+\mathrm{IIE}(\textcolor{red}{1}\boldsymbol\cdot)=\mathrm{IIE}(\textcolor{blue}{0}\boldsymbol\cdot)+\mathrm{IDE}(\boldsymbol{\cdot}\textcolor{red}{1})$. This sum, termed the \textit{overall effect} (OE) by \cite{VanderWeele2014a}, is the effect of shifting the exposure from \textcolor{blue}{0} to \textcolor{red}{1} and shifting the mediator distribution from $d_{\textcolor{blue}{M(0)}|C}$ to $d_{\textcolor{red}{M(1)}|C}$. The total effect, on the other hand, is the effect of shifting exposure from \textcolor{blue}{0} to \textcolor{red}{1} without intervening on the mediator.
Second, in the special case with no intermediate confounders (Fig. \ref{fig:NIDE}a), the interventional (in)direct effects are equal to their natural counterparts (and OE is equal to TE); in the general case (Fig. \ref{fig:SIDE}a), these effects are generally not the same.%
\footnote{In the special case, the equality of mean potential outcomes, $\E[Y(a',M(a))]=\E[Y(a',\mathcal{M}(a|C)$, follows from the fact that $M(a)$ and $\mathcal{M}(a|C)$ share the same distribution. In the general case (Fig. \ref{fig:SIDE}), $L$ is a mediator that precedes $M$, and $Y$ depends on both mediators. Thus $Y(a',\mathcal{M}(a|C))$ is shorthand for $Y(a',L(a'),\mathcal{M}(a|C))$ (where the first mediator follows naturally after exposure condition $a'$, taking its potential value $L(a')$, and the second mediator is a random draw from $d_{M(a)|C}$); and $Y(a',M(a))$ is shorthand for $Y(a',L(a'),M(a))$. Now we do not have the same equality of mean potential outcomes, because the joint distribution of $\{(L(a'),\mathcal{M}(a|C)\}$ is different from the joint distribution of $\{L(a'),M(a)\}$ -- conditional on $C$, $L(a')$ and $\mathcal{M}(a|C)$ are independent, but $L(a')$ and $M(a)$ are generally dependent.}
Third, there is a difference in \textit{identification} (which we comment on further in the last section of the paper): natural effects identification requires the no intermediate confounders (no $L$) assumption; interventional effects identification does not. 
Based on the last two points, a practical takeaway is: (i) in the special case with no $L$, it does not matter whether we are interested in natural or interventional (in)direct effects (or both) because they coincide; (ii) in the general case with $L$ variables, these effects do not coincide, and natural effects are not identified while interventional effects may be (if their identification assumptions hold).

A side note: For an explanation of the difference in identification between the two effect types (which is tangential to our current focus on connecting effect definitions to research questions), see \cite{VanderWeele2014a}. A numerical example of these effects (which references identification) is provided in the Supplementary Material; a similar example concerning natural effects can be found in \cite{Pearl2012}.

\medskip

\noindent\textbf{Interventional (in)direct effects in the college prep program example.}
Equipped with the interventional (in)direct effects, let us now revisit the questions asked by the investigators of the college prep program: what would be the effect of the program (i) if program components that only serve to improve self-awareness were eliminated, (ii) if only self-awareness-related components were kept, or (iii) if some other modification were made.
Let it be clear upfront that the interventional (in)direct effects do not exactly answer these questions, or any other questions about realistic program modifications. Rather, they concern \textit{ideal} interventions that intervene on the exposure and the mediator distribution without changing anything else in the causal structure -- they do not have additional effects on any other variables, and do not affect how variables influence one another.%
\footnote{For the technical details of these requirements, we refer the reader to \citet{Lok2016}.}
It is unlikely that any real world program modification qualifies as such an ideal intervention. Where there is an approximate correspondence, however, the ideal intervention effect provides a sense about what a realistic intervention effect might be; how good the approximation is depends on how close the realistic intervention is to the ideal intervention.

For the first of the three questions above, $\mathrm{IDE}(\boldsymbol{\cdot}\textcolor{blue}{0})$ is relevant. $\mathrm{IDE}(\boldsymbol{\cdot}\textcolor{blue}{0})$ contrasts two interventions: one setting exposure to \textcolor{red}{1}, the other setting exposure to \textcolor{blue}{0}, both setting mediator distribution to $d_{\textcolor{blue}{M(0)}|C}$. Let's call these the active intervention condition and the comparison intervention condition. The active intervention condition corresponds approximately to the modified program without self-awareness promoting components. The modified program is desired to achieve the same mediator distribution as $d_{\textcolor{blue}{M(0)}|C}$, but not expected to result in each individual having their specific $\textcolor{blue}{M(0)}$ value. The comparison intervention condition can be thought of as corresponding to a modified control condition -- which, similar to the modified program, might achieve the same mediator distribution but not the same individual specific values. This is relevant, for example, if our research uses the equal attention control strategy, so the control condition for the modified program might be shorter than the control condition for the original program in the current study. (We comment shortly on the case where the control condition is unchanged.)
Similarly, $\mathrm{IIE}(\textcolor{blue}{0}\boldsymbol\cdot)$ is relevant to the investigators' second question. The active intervention condition in $\mathrm{IIE}(\textcolor{blue}{0}\boldsymbol\cdot)$ sets exposure to \textcolor{blue}{0} and mediator distribution to $d_{\textcolor{red}{M(1)}|C}$, which corresponds approximately to the modified program retaining only self-awareness related components. 
The comparison intervention condition is the same as that in $\mathrm{IDE}(\boldsymbol\cdot\textcolor{blue}{0})$, discussed above. 
As for the investigators' third question, the interventional (in)direct effects are clearly not relevant, due to lack of correspondence with the program modification in question.

A couple of comments are warranted. First, an analysis targeting interventional effects (motivated by \textit{what if} questions) looks different from an analysis targeting natural effects (motivated by the desire to decompose the total effect). A TE-decomposing analysis invariably involves identifying and estimating a pair (or two pairs) of direct and direct effects. A \textit{what if} analysis is not concerned with pairs of effects; rather, it targets the specific effect (or effects) most relevant to the substantive research question. Defining effects based on the conditions we wish to contrast, in our opinion, is the gist of the interventional effects approach.

Second, depending on the specific study and the specific question, the interventional (in)direct effects may or may not provide the best approximation to the real world contrast of interest to the researcher. We had a glimpse of this above, where the comparison condition in $\mathrm{IDE}(\boldsymbol{\cdot}\textcolor{blue}{0})$ and $\mathrm{IIE}(\textcolor{blue}{0}\boldsymbol{\cdot})$ doesn't quite fit if we imagine wanting to use the same control condition as in the current study (e.g., the same equal attention control condition, or the same no engagement condition) in evaluating the modified program. Also, all program modifications that fit in the investigators' third question are completely off limits to our gaze through the lense of interventional (in)direct effects. Fortunately, there is some rectification of these limits using the general class of interventional effects.


\medskip

\subsection{Interventional effects more generally}

\noindent Interventional (in)direct effects are one special type within the broader class of interventional effects; the special feature is that they are direct and indirect effects. The general class is much larger. OE, for example, is an interventional effect as it contrasts two intervention conditions. TE is also an interventional effect, contrasting an intervention that sets exposure to \textcolor{red}{1} and an intervention that sets exposure to \textcolor{blue}{0}. Any contrast of what happens between two different intervention conditions, or between an intervention condition and no intervention, belongs in the class of interventional effects.

Tapping into this general class of effects allows researchers to define effects that correspond better to their real world questions. In the college prep program example, if the control condition is unchanged, then the interventional effects $\E[Y(\textcolor{red}{1},\mathcal{M}(\textcolor{blue}{0}|C))]\!-\!\E[\textcolor{blue}{Y(0)}]$ and $\E[Y(\textcolor{blue}{0},\mathcal{M}(\textcolor{red}{1}|C))]\!-\!\E[\textcolor{blue}{Y(0)}]$ are relevant to the investigators' first and second questions, respectively.%
\footnote{$\E[Y(\textcolor{red}{1},\mathcal{M}(\textcolor{blue}{0}|C))]\!-\!\E[\textcolor{blue}{Y(0)}]$ is not a direct effect. It is the combination of (a) the effect of shifting the mediator from the natural $\textcolor{blue}{M(0)}$ to the randomly drawn $\mathcal{M}(\textcolor{blue}{0}|C)$, and (b) the direct effect $\mathrm{IDE}(\boldsymbol{\cdot}\textcolor{blue}{0})$. Similarly, $\E[Y(\textcolor{blue}{0},\mathcal{M}(\textcolor{red}{1}|C))]\!-\!\E[\textcolor{blue}{Y(0)}]$ is not an indirect effect; it combines (a) and the indirect effect $\mathrm{IIE}(\textcolor{blue}{0}\boldsymbol\cdot)$.}
We now use a different example to illustrate that flexible application of interventional effects helps address a wide range of \textit{what if} questions.

This example draws from disparities research. The population is adolescents. The exposure (or group) variable is sexual minority status, $A=1$ if the individual identifies as lesbian, gay, bisexual or another sexual minority identity, $A=0$ otherwise. The outcome, $Y$, is any measure of well-being or lack thereof (e.g., life satisfaction, depressive symptoms). The mediator of interest, $M$, is experience of bullying in school. Let $C$ denote demographic/context variables that are unlikely to be influenced by sexual minority status (e.g., age, sex, anti-discrimination and same-sex legislation, geographical political leaning, economic climate, etc.) that may influence well-being or bullying experience. Relative to sexual majority (heterosexual) adolescents with similar $C$ values, sexual minority adolescents tend to experience more bullying, $\E[M|A=\textcolor{red}{1},C]>\E[M|A=\textcolor{blue}{0},C]$. Sexual minority adolescents also tend to be lower on well-being measures. Within levels of $C$, the well-being disparity associated with sexual minority status is 
$$\text{disparity}(C)=\E[Y|A=\textcolor{red}{1},C]-\E[Y|A=\textcolor{blue}{0},C],$$
which, averaged over the distribution of $C$ in the sexual minority adolescent subpopulation gives the population disparity measure. To avoid unnecessarily complicating the argument, we simply consider the conditional measure $\text{disparity}(C)$. This disparity measure is analogous to the total effect previously constructed, and it would turn into a total effect if we were willing to inject a few additional inputs in our construction.%
\footnote{Those inputs are: (i) admission of the idea that $\textcolor{blue}{Y(0)}$ is well defined for sexual minority adolescents, and (ii) assumption that given $C$, sexual minority/majority status is independent of $\textcolor{blue}{Y(0)}$. These would imply that $\text{disparity}(C)$ defined above is equal to $\E[\textcolor{red}{Y(1)}|A=1,C]-\E[\textcolor{blue}{Y(0)}|A=1,C]$, a total causal effect on sexual minority adolescents. There is a long-standing debate about (i) that is important but tangential to our current discussion, and (ii) is a very strong assumption.}
Here we do not need to construct a total effect in the process of defining the interventional effects of interest.

In this example, the investigators ask two questions. The first question is completely hypothetical: \textit{How much of the disparity in well-being would be removed if we could reduce the level of bullying experienced by sexual minority adolescents down to the level experienced by sexual majority adolescents?} In the language of interventional effects, this translates to swapping a sexual minority adolescent's natural bullying experience value for a value randomly drawn from the distribution of bullying experience of sexual majority adolescents with the same $C$ value. Denote this distribution by $d_{M|\textcolor{blue}{0},C}$ and the random draw by $\mathcal{M}_{|\textcolor{blue}{0},C}$. Using this intervention condition, we split this disparity into two parts:
\begin{align*}
    &\text{dis}\text{parity}(C)=
    \\
    &~~~~~\,
    \underbrace{\E[Y\mid A\!=\!\textcolor{red}{1},\!C]-\E[Y(\textcolor{red}{1},\mathcal{M}_{|\textcolor{blue}{0},C})\mid A\!=\!\textcolor{red}{1},\!C]}_{\textstyle\text{disparity-removed}(C)}+
    \underbrace{\E[Y(\textcolor{red}{1},\mathcal{M}_{|\textcolor{blue}{0},C})\mid A\!=\!\textcolor{red}{1},\!C]-\E[Y\mid A\!=\!\textcolor{blue}{0},\!C]}_{\textstyle\text{remaining-disparity}(C)}.
\end{align*}
Here \textit{disparity removed} is an interventional effect on the sexual minority subpopulation, contrasting their well-being (i) under the intervention condition that sets the mediator (bullying experience) distribution to $d_{M|\textcolor{blue}{0},C}$ and (ii) under the no intervention condition. It represents the improvement in well-being for sexual minority adolescents as a result of this hypothetical intervention. It may be interesting to note that \textit{remaining disparity}, like the original disparity measure, is an across-group contrast (i.e., a difference in well-being between sexual minority and sexual majority adolescents), and thus is not an interventional (or causal) effect. The key distinction is that causal effects are defined as contrasts of different potential outcomes for the same group or the same individual.

Up to this point, this example, although involving a different context (mixing disparities and interventional effects, and considering effects on the exposed subpopulation instead of the full population), is similar to the college prep program example in that the natural mediator distribution under one exposure condition is replaced by the mediator distribution from the other exposure condition. The next question breaks this pattern.

Suppose the investigators work with a school board that is considering adopting an anti-bullying intervention that is expected to reduce the bullying experienced by sexual minority adolescents but not to the level of equating it to that experienced by sexual majority students. They ask: \textit{What would be the improvement of well-being for sexual minority adolescents if this intervention could reduce their experience of bullying in school down to halfway between the current levels of the two groups?}
Now with this intervention, the bullying experience of a sexual minority adolescent would not come from $d_{M|\textcolor{red}{1},C}$ or $d_{M|\textcolor{blue}{0},C}$, but a mixture of these two distributions. Denoting this (assumed half-half) mixture by $d_{M|\textcolor{violet}{0.5},C}$, and a random draw from it by $\mathcal{M}_{|\textcolor{violet}{0.5},C}$, we have the intervention effect
$$\E[Y|A=\textcolor{red}{1},C]-\E[Y(\textcolor{red}{1},\mathcal{M}_{|\textcolor{violet}{0.5},C})|A=\textcolor{red}{1},C],$$

\noindent which is the (approximate) improvement in the well-being of sexual minority adolescents to be expected as a result of the anti-bullying intervention. (This effect is not labeled \textit{disparity removed} here, because the intervention may also benefit sexual majority adolescents.)

\smallskip

The two examples above give a glimpse of the flexible applicability of interventional effects, all based on the simple idea: what intervention conditions do we wish to contrast? The common thread is that these are (ideal) interventions that set variables to specific values, or set their distributions to specific distributions, that are \textit{priorly determined}.%
\footnote{The pair of adjectives \textit{deterministic} and \textit{stochastic} are used to differentiate a fixed value and a random value. An intervention that sets a variable to a fixed value (e.g., setting exposure to 1) is a deterministic intervention; one that sets the distribution of a variable (e.g., setting mediator distribution to $d_{M|\textcolor{violet}{0.5},W}$) is a stochatic intervention.}
In the examples the mediator distribution is set to $d_{\textcolor{red}{M(1)}|C}$, $d_{\textcolor{blue}{M(0)}|C}$ or a mixture of these, but any other distribution that suits the research question can be used. The strategy of setting the distribution of a variable has also been applied to exposure variables \citep{Diaz2019,Kennedy2018} in some other settings. Our recommendation, if the researcher is approaching a mediation setting with an interventional instead of an explanatory perspective, is to flexibly define interventional effects based on the specific \textit{what if} questions, and not simply default to the IDEs and IIEs.

\smallskip

One more comment before we close this section: the broad class of interventional effects includes a special set of effects \citep[defined by][]{Didelez2006,Geneletti2007} that we refer to as \textit{generalized} interventional direct effects (GIDE).%
\footnote{These are referred to in \cite{Didelez2006} as \textit{standardized direct effects}, based on the concept of population standardization in demography. \cite{Geneletti2007} uses the term \textit{generated direct effects}.}
GIDEs are similar to IDEs, except rather than only two choices for the mediator distribution (which give the two effects $\mathrm{IDE}(\boldsymbol{\cdot}\textcolor{blue}{0})$ and $\mathrm{IDE}(\boldsymbol{\cdot}\textcolor{red}{1})$), we can use any reasonable distribution of choice for the mediator. For a chosen mediator distribution $\mathscrsfs{D}$, we have
$$\mathrm{GIDE}(\boldsymbol{\cdot}\mathscrsfs{D})=\E[Y(\textcolor{red}{1},\mathcal{M}_\mathscrsfs{D})]-\E[Y(\textcolor{blue}{0},\mathcal{M}_\mathscrsfs{D})],$$
where $\mathcal{M}_\mathscrsfs{D}$ is a random draw from the distribution $\mathscrsfs{D}$. While IDEs are paired with IIEs, GIDEs (that are not IDEs) are not paired with indirect effects.%
\footnote{\cite{Geneletti2007} defines an indirect effect as the difference between TE and a direct effect, so based on that definition, a direct effect is always paired with an indirect effect. While TE minus any direct effect might be an interesting mathematical object, it seems calling it an indirect effect not meaningful unless the direct effect is an NDE. We maintain the seemingly common view -- reflected in the definition of NIEs and IIEs and in the common assertion that controlled direct effects are not paired with indirect effects -- that the term \textit{indirect effect} should be reserved to refer to effects of \textit{shifts in the mediator value or distribution} that are as if \textit{in response to a switch of the exposure from 0 to 1}.}
We will comment on the potential relevance of GIDEs after introducing the one last effect type in this paper.

\subsection{Controlled direct effects}

\begin{figure*}[h!]
\caption{Controlled direct effect}\label{fig:CDE}
    \centering
\resizebox{.8\textwidth}{!}{%
\begin{tikzpicture}[
obs/.style={rectangle, minimum size=5mm},
unobs/.style={rectangle, minimum size=4mm},
setred/.style={rectangle, draw=red, thick, minimum size=5mm},
setblue/.style={rectangle, draw=blue, thick, minimum size=5mm},
setblack/.style={rectangle, draw=black, thick, minimum size=5mm}
]
\node[obs]    (C)                            {$C$};
\node[obs]    (A)  [right=of C, xshift=+1mm] {$A$};
\node[obs]    (M)  [right=of A, xshift=+1mm] {$M$};
\node[obs]    (Y)  [right=of M, xshift=+1mm] {$Y$};
\node[unobs]  (UA) [below=of A, yshift=+7mm] {\footnotesize $U_A$};
\node[unobs]  (UM) [below=of M, yshift=+7mm] {\footnotesize $U_M$};
\node[unobs]  (UY) [below=of Y, yshift=+7mm] {\footnotesize $U_Y$};

\draw[->] (C) -- (A);
\draw[->] (C) .. controls +(up:15mm) and +(up:15mm)  .. (Y);
\draw[->] (C) .. controls +(up:10mm) and +(up:10mm)  .. (M);
\draw[thick, ->] (A) -- (M);
\draw[thick, ->] (M) -- (Y);
\draw[->]        (UA) -- (A);
\draw[->]        (UM) -- (M);
\draw[->]        (UY) -- (Y);
\draw[thick, ->] (A) .. controls +(up:10mm) and +(up:10mm)  .. (Y);

\node[obs]     (C10)  [right=of Y, xshift=+10mm]   {$C$};
\node[setred]  (A10)  [right=of C10, xshift=-1mm] {$A=\textcolor{red}{1}$};
\node[setblack] (M10)  [right=of A10, xshift=-1mm] {$M=m$};
\node[obs]     (Y10)  [right=of M10]              {$Y(\textcolor{red}{1},m)$};
\node[unobs]   (UY10) [below=of Y10, yshift=+7mm] {\footnotesize $U_Y$};

\draw[thick, ->] (M10) -- (Y10);
\draw[->]        (UY10) -- (Y10);
\draw[->] (C10) .. controls +(up:15mm) and +(up:15mm)  .. (Y10);
\draw[thick, ->] (A10) .. controls +(up:10mm) and +(up:10mm)  .. (Y10);

\node[obs]     (C01)  [below=of C10, yshift=-12mm] {$C$};
\node[setblue] (A01)  [right=of C01, xshift=-1mm]  {$A=\textcolor{blue}{0}$};
\node[setblack]  (M01)  [right=of A01, xshift=-1mm]  {$M=m$};
\node[obs]     (Y01)  [right=of M01]               {$Y(\textcolor{blue}{0},m)$};
\node[unobs]   (UY01) [below=of Y01, yshift=+7mm]  {\footnotesize $U_Y$};

\draw[thick, ->] (M01) -- (Y01);
\draw[->]        (UY01) -- (Y01);
\draw[->] (C01) .. controls +(up:15mm) and +(up:15mm)  .. (Y01);
\draw[thick, ->] (A01) .. controls +(up:10mm) and +(up:10mm)  .. (Y01);

\node[obs] [above=of A, yshift=+7mm, xshift=+9mm] {a. The regular world};
\node[obs] [above=of M10, yshift=+3mm, xshift=-9mm] {\begin{tabular}{c}b. The two worlds contrasted in \\ the direct effect controlling mediator at level $m$\end{tabular}};
\end{tikzpicture}%
}
\end{figure*}
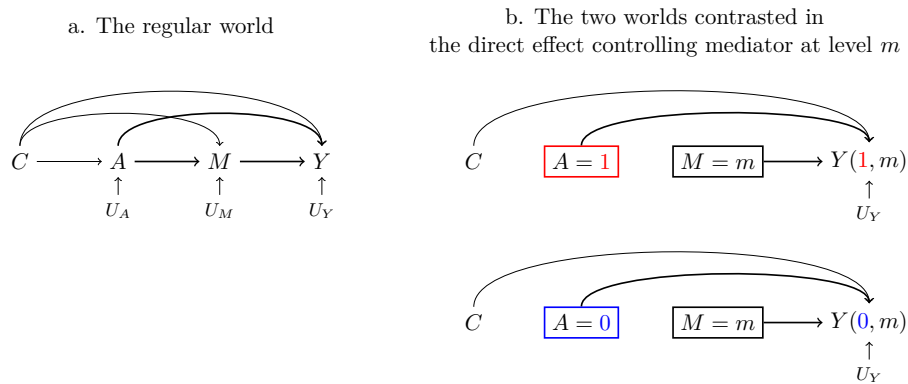

\noindent \textit{Controlled direct effects}, the oldest among all the (in)direct effects in this paper, are effects of the exposure if the mediator were controlled, i.e., set to a specific level for everyone. For mediator control level $m$, the \textit{individual controlled direct effect} is defined as
$${CDE}_i(m)=Y_i(\textcolor{red}{1},m)-Y_i(\textcolor{blue}{0},m),$$
i.e., the contrast between the potential outcomes for individual $i$ if their exposure is respectively set to the exposed and unexposed conditions, and their mediator is set to $m$ (see Fig. \ref{fig:CDE}). Controlled direct effects may vary across individuals, and within an individual may vary depending on the mediator control value $m$. The \textit{average controlled direct effect} for mediator control level $m$ is the average of the corresponding individual effects,
$$\mathrm{CDE}(m)=\E[Y(\textcolor{red}{1},m)]-\E[Y(\textcolor{blue}{0},m)].$$
CDEs are not paired with indirect effects. A CDE is an effect of the exposure when controlling the mediator. One could imagine an effect of the mediator when controlling the exposure, but that is not in any sense an indirect effect of the exposure.

\medskip

\begin{table*}[h!]
\caption{A summary of the effects}\label{tab:summary}
\resizebox{\textwidth}{!}{%
{\footnotesize
\begin{tabular}{>{\raggedright}p{7.5cm}>{\raggedright\arraybackslash}p{10cm}}
    \hline
    \\[-.5em]
    \multicolumn{2}{@{}l}{\textbf{Natural (in)direct effects -- explaining the total effect}}\smallskip
    \\[.5em]
    \multicolumn{1}{@{}l}{\textsc{Total effect decompositions}} & \multicolumn{1}{@{}l}{\textsc{Relevant research questions}}\medskip
    \\
    direct-indirect:~~~$\mathrm{TE}=\mathrm{NDE}(\boldsymbol\cdot\textcolor{blue}{0})+\mathrm{NIE}(\textcolor{red}{1}\boldsymbol\cdot)$ & Does the effect of the college prep program include an indirect (mediated by self awareness) component?\smallskip
    \\
    indirect-direct:~~~$\mathrm{TE}=\mathrm{NIE}(\textcolor{blue}{0}\boldsymbol\cdot)+\mathrm{NDE}(\boldsymbol\cdot\textcolor{red}{1})$ & Does the effect of the college prep program include a direct (not mediated by self awareness) component?\smallskip
    \\
    both & What can we learn about the effect of the college prep program, either through self awareness or through other mechanisms?\smallskip
    \\\hline
    \\[-.5em]
    \multicolumn{2}{@{}l}{\textbf{Interventional effects -- effects of hypothetically modified exposures or hypothetical interventions}}\smallskip
    \\[.5em]
    \multicolumn{1}{@{}l}{\textsc{Several special effect types}} & \multicolumn{1}{@{}l}{\textsc{Links to other special effect types}} \medskip
    \\
    interventional direct effects (IDEs) & paired with IIEs; special case of GIDEs\smallskip
    \\
    interventional indirect effects (IIEs) & paired with IDEs\smallskip
    \\
    controlled direct effects (CDEs) & special case of GIDEs\smallskip
    \\
    generalized interventional direct effects (GIDEs) & contain IDEs and CDEs as special cases\smallskip
    \\
    overall interventional effect (OE) & sum of $\mathrm{IDE}(\boldsymbol\cdot\textcolor{blue}{0})$ and $\mathrm{IIE}(\textcolor{red}{1}\boldsymbol\cdot)$; sum of $\mathrm{IIE}(\textcolor{blue}{0}\boldsymbol\cdot)$ and $\mathrm{IDE}(\boldsymbol\cdot\textcolor{red}{1})$\smallskip
    \\
    total effect (TE) & decomposed by natural (in)direct effects\smallskip
    \\
    \\
    \multicolumn{1}{@{}l}{\textsc{More generally,}}
    \\
    \multicolumn{1}{@{}l}{\textsc{variation in conditions contrasted}} & \multicolumn{1}{@{}l}{\textsc 
    {Examples}}\medskip
    \\
    set exposure distribution to $\mathscrsfs D_A$ and mediator distribution to $\mathscrsfs D_M$ (most general intervention) & include all the examples below\bigskip\bigskip
    \\
    set exposure to $a$ and mediator distribution to $\mathscrsfs D$ (relatively general intervention) & an anti-bullying program that brings the bullying experienced by sexual minority adolescents down to halfway between sexual minority and majority levels\smallskip
    \\
    & the city continuing the injury prevention program plus implementing the city-wide intervention setting home water heating systems to 100-120$^\circ$F\medskip
    \\
    set exposure to $a$ and mediator distribution to $d_{M(a')|C}$ (specific intervention) & a hypothetical intervention that brings the bullying experienced by sexual minority adolescents down to the level experienced by sexual majority adolescents\smallskip
    \\
    & modified college prep program without self-awareness components\smallskip
    \\
    & modified college prep program with only self-awareness components\medskip
    \\
    set exposure to $a$ (specific intervention) & college prep program\smallskip
    \\
    & a control condition with some engagement (or a placebo)\medskip
    \\
    set exposure to $a$ and mediator to $m$ (very specific intervention) & the city continuing the injury prevention program plus implementing the structural intervention setting home water heating systems to 120$^\circ$F\medskip
    \\
    & the city discontinuing the injury prevention program, but implementing the structural intervention setting home water heating systems to 120$^\circ$F\smallskip
    \\
    no intervention & a no engagement control condition\smallskip
    \\
    & simply observing the bullying experience of sexual minority adolescents\smallskip
    \\\hline
\end{tabular}%
}}
\end{table*}

\noindent\textbf{When are CDEs of interest?} CDEs are a type of interventional effect. If a mediator level $m$ is desirable for all and if a feasible and ethical way exists to set it for all, then $\mathrm{CDE}(m)$ is relevant. Interventions that shift and fix a variable for a whole population are generally structural, e.g., seatbelt laws, age limits for alcohol/cigarette sale, city-wide water treatment, etc. Consider one last simple example: A city has a successful childhood injury prevention program, and researchers have figured out that a mechanism of the program's success is that it increased parents' awareness of burn risks and knowledge of how to set the water heater in their homes to a safe temperature, which shifted the water temperature down on average in households that participated in the program, leading to fewer burns in small children. Recognizing this important effect, the city has passed a bill setting a legal home water heating temperature of maximum 120$^{\circ}$F and is planning a blanket intervention sending technicians door-to-door to help set water heaters to under this temperature. The city wants to know whether their childhood injury prevention program would still be effective in the new reality where the burns-due-to-hot-water problem has been taken care of. They are interested in $\mathrm{CDE}(\text{water temperature = 120})$. 

\smallskip

CDEs (at the population average level) are a special type of GIDEs, where the distribution $\mathscrsfs{D}$ is a single point $m$. This means that when we are interested in a $\mathrm{CDE}(m)$ where $m$ is the desired control mediator level, but suspect that what is obtained would be a range of values that may be centered at $m$ but with some variation, we can simply switch to $\mathrm{GIDE}(\boldsymbol\cdot\mathscrsfs{D})$ where the distribution $\mathscrsfs{D}$ is defined to reflect that variation.

\medskip

This closes the effect definition topic. A summary of the different effect types presented is given in Table \ref{tab:summary}.

\section{Closing Remarks}

\noindent The focus of this paper has been  the first step in analysis, selecting the target causal effect(s) that reflect the substantive research question. In place of conclusions, we offer a few comments on the next steps -- effect identification and estimation -- and on settings that are more complex.

\medskip

\noindent\textbf{Effect identification.} Identification of an effect depends on identification of the mean outcome for each of the two contrasted conditions. Our first comment, as an orientation for readers who are not familiar with this topic, is that there is a hierarchy of difficulty for identification of the mean outcome for different conditions. Identification of the mean outcome for the \textit{no manipulation} condition is the easiest; it is simply the mean observed outcome. Mean outcome identification for conditions where \textit{exposure is set to one value} and everything follows naturally (both conditions in TE) is harder, requiring the assumption that exposure assignment is independent of this outcome, possibly given observed pre-exposure covariates $C$, usually referred to as no unobserved $A$-$Y$ confounding. Next up the ladder of difficulty are conditions where \textit{exposure is set to one value} and \textit{mediator is set to one value or a known distribution} (both conditions in a CDE), where identification requires an additional assumption: no unobserved $M$-$Y$ confounding given $C,A$ and intermediate confounders $L$ (if any). If the known distribution is replaced with $d_{M(a)|C}$ (both conditions in each IDE/IIE), an additional assumption is needed: no unobserved $A$-$M$ confounding given $C$. At the top of the difficulty ladder are conditions where \textit{exposure is set to one value but mediator is set to its natural value under the other exposure condition} (the in-between world in each NDE/NIE), where identification requires that the no unobserved $M$-$Y$ confounding assumption holds conditional on $C$ only, that is, no intermediate confounders are allowed.%
\footnote{This is also referred to as the \textit{cross-world} independence assumption, because formally it is $M(a)\independent Y(a',m)\mid C,A$, and for $a\neq a'$, variables $M(a)$ and $Y(a',m)$ do not live in the same world.}
This is a rough summary of a key part of the identification picture. There are details within each of these assumptions that we necessarily gloss over, and there are other assumptions including no interference, consistency, composition, and positivity. We refer the reader to the rich literature, e.g., \cite{Pearl2001,Petersen2006,Imai2010a,Imai2010,VanderWeele2009,Didelez2006,VanderWeele2014a,Pearl2012,Avin2005}.

Our second comment concerns the case where the natural (in)direct effects are not identified (due to the presence of intermediate confounders $L$) but interventional counterparts are. In this challenging case, it may be tempting to simply estimate the latter, even when the scientific interest is in explaining the total effect, that is, in the natural effects. We recommend caution here. \cite{VanderWeele2017} pointed out that except in rare cases, when an IIE/IDE is non-zero, its natural counterpart is generally also non-zero, which we take to mean that we might consider using the IIE/IDE as proxy for the purpose of testing whether the NIE/NDE is zero. If we are interested in the magnitude of the natural effects that decompose TE, however, the interventional effects are a suboptimal approximation; there is an alternative strategy -- see explanations in footnote.%
\footnote{To be concrete, take a natural effect that is a contrast of $\E[Y(a,M(a'))]$ where $a\neq a'$ and $\E[Y(a)]$, while the corresponding interventional effect contrasts $\E[Y(a,\mathcal{M}(a'|C))]$ and $\E[Y(a,\mathcal{M}(a|C))]$. First, under the given identification assumptions, $\E[Y(a)]$ is identified, so there is no obvious reason to approximate it with $\E[Y(a,\mathcal{M}(a|C))]$ that is both incorrect and more complicated to estimate. Second, the reason $\E[Y(a,M(a'))]=\E[Y(a,L(a),M(a'))]$ is unidentified is that $L(a)$ and $M(a')$ are dependent, even after conditioning on $C$. Approximation by $\E[Y(a,\mathcal{M}(a'|C))]=\E[Y(a,L(a),\mathcal{M}(a'|C))]$, however, is equivalent to assuming away this dependence. An alternative that respects this dependence is to assume a more reasonable level or consider a range for it (and obtain bounds for the natural effects). One could assume this dependence is bounded below by independence and above by the same-world dependence of $L$ and $M$. A strategy similar to this was used in \cite{Daniel2015}.}
We also recommend, if one ever uses IDE/IIEs to approximate NDE/NIEs, to be explicit about this approximation, so that there is no ambiguity in the interpretation of analysis results.

\medskip

\noindent\textbf{Effect estimation.} There is a huge and fast growing literature on estimation, and our comments here are limited. Estimation in the mediation setting can be considered an extension of the non-mediation setting, where estimation of the effect of an exposure on an outcome may rely on an outcome model (e.g., when using regression), or an exposure assignment model (e.g., when using propensity score matching or weighting), or both models (for double robustness). In the mediation setting, if there are no intermediate confounders, estimation of (in)direct effects, and more generally, estimation of effects of interventions that intervene on both the exposure and the mediator, may rely on a combination of two out of three models (a model for the outcome mean, a model for the mediator distribution, and a model for exposure assignment), or all three models for robustness (robustness requires any two of the three models to be correct). With an intermediate confounder, an additional model is required. The methods may involve regression, weighting and imputation; and the term regression is used in a general sense to include both parametric and semiparametric models. We refer the reader to the rich literature, e.g., \cite{Imai2010a,Imai2010,Pearl2012,Valeri2013,Rudolph2017,Tingley2014,VanderWeele2010,Daniel2015,Hong2015,Hong2010,VanderWeele2013a,Lange2012,Vansteelandt2012,Vansteelandt2012a,TchetgenTchetgen2012}.

The models used in traditional mediation analysis are the closest to the estimation strategy that relies on modeling the mediator and the outcome. Two features of traditional mediation analysis deviate from this estimation strategy of causal mediation analysis. First, the outcome model in traditional mediation analysis rarely includes an $A$-$M$ interaction; causal mediation analysis generally does not impose this restriction. Second, after fitting models, traditional mediation analysis computes the product of coefficients, while causal mediation analysis computes target causal effects. As previously mentioned, these results line up only in special cases.

Since identification requires untestable assumptions -- mediators, unlike exposures, cannot be randomized -- sensitivity analyses are recommended after, or as part of, effect estimation. See e.g., \cite{Imai2010,Imai2010a,Ding2016} for sensitivity analyses for unobserved mediator-outcome confounding.

\medskip

\noindent\textbf{More complex settings.} Methods for causal mediation analysis -- spanning effect definition, identification, and estimation -- of more complex settings is very much an active area of research. We refer the reader to relevant literature. For causal mediation analysis with survival data, see e.g., \cite{Didelez2018,Lange2011,Valeri2015,Vanderweele2011,TchetgenTchetgen2011}. For multiple mediators, see e.g., \cite{VanderWeele2013a,Daniel2015,Vansteelandt2017}. For time-varying mediators and exposures, see e.g., \cite{Zheng2017,VanderWeele2017}.

\bigskip

\noindent
To sum up, this paper discusses a key decision in mediation analysis, the selection of the estimand. The proposal is simple and obvious: try to best match the estimand to the research question. The rest of the analysis -- including technical aspects of identification and estimation, and the interpretation and discussion of results -- resolves around the selected estimand. The causal contrasts presented in the paper, including natural (in)direct effects and a wide range of interventional effects, offer the researcher flexibility, and conceptual clarity, in targeting the knowledge they desire. 

\bibliography{refs}

\begin{thebibliography}{}

\bibitem [\protect \citeauthoryear {%
Avin%
, Shpitser%
\BCBL {}\ \BBA {} Pearl%
}{%
Avin%
\ \protect \BOthers {.}}{%
{\protect \APACyear {2005}}%
}]{%
Avin2005}
\APACinsertmetastar {%
Avin2005}%
\begin{APACrefauthors}%
Avin, C.%
, Shpitser, I.%
\BCBL {}\ \BBA {} Pearl, J.%
\end{APACrefauthors}%
\unskip\
\newblock
\APACrefYearMonthDay{2005}{}{}.
\newblock
{\BBOQ}\APACrefatitle {{Identifiability of Path-Specific Effects}}
  {{Identifiability of Path-Specific Effects}}.{\BBCQ}
\newblock
\APACjournalVolNumPages{Proceedings of the International Joint Conference on
  Artificial Intelligence, Edinburgh, Scotland}{}{}{357--363}.
\PrintBackRefs{\CurrentBib}

\bibitem [\protect \citeauthoryear {%
Baron%
\ \BBA {} Kenny%
}{%
Baron%
\ \BBA {} Kenny%
}{%
{\protect \APACyear {1986}}%
}]{%
Baron1986}
\APACinsertmetastar {%
Baron1986}%
\begin{APACrefauthors}%
Baron, R\BPBI M.%
\BCBT {}\ \BBA {} Kenny, D\BPBI A.%
\end{APACrefauthors}%
\unskip\
\newblock
\APACrefYearMonthDay{1986}{}{}.
\newblock
{\BBOQ}\APACrefatitle {{The moderator-mediator variable distinction in social
  psychological research: conceptual, strategic, and statistical
  considerations.}} {{The moderator-mediator variable distinction in social
  psychological research: conceptual, strategic, and statistical
  considerations.}}{\BBCQ}
\newblock
\APACjournalVolNumPages{Journal of Personality and Social
  Psychology}{51}{6}{1173--1182}.
\newblock
\begin{APACrefDOI} \doi{10.1037/0022-3514.51.6.1173} \end{APACrefDOI}
\PrintBackRefs{\CurrentBib}

\bibitem [\protect \citeauthoryear {%
Daniel%
, {De Stavola}%
, Cousens%
\BCBL {}\ \BBA {} Vansteelandt%
}{%
Daniel%
\ \protect \BOthers {.}}{%
{\protect \APACyear {2015}}%
}]{%
Daniel2015}
\APACinsertmetastar {%
Daniel2015}%
\begin{APACrefauthors}%
Daniel, R.%
, {De Stavola}, B\BPBI L.%
, Cousens, S\BPBI N.%
\BCBL {}\ \BBA {} Vansteelandt, S.%
\end{APACrefauthors}%
\unskip\
\newblock
\APACrefYearMonthDay{2015}{}{}.
\newblock
{\BBOQ}\APACrefatitle {{Causal mediation analysis with multiple mediators}}
  {{Causal mediation analysis with multiple mediators}}.{\BBCQ}
\newblock
\APACjournalVolNumPages{Biometrics}{71}{}{1--14}.
\newblock
\begin{APACrefDOI} \doi{10.1111/biom.12248} \end{APACrefDOI}
\PrintBackRefs{\CurrentBib}

\bibitem [\protect \citeauthoryear {%
D{\'{i}}az%
\ \BBA {} Hejazi%
}{%
D{\'{i}}az%
\ \BBA {} Hejazi%
}{%
{\protect \APACyear {2019}}%
}]{%
Diaz2019}
\APACinsertmetastar {%
Diaz2019}%
\begin{APACrefauthors}%
D{\'{i}}az, I.%
\BCBT {}\ \BBA {} Hejazi, N\BPBI S.%
\end{APACrefauthors}%
\unskip\
\newblock
\APACrefYearMonthDay{2019}{}{}.
\newblock
{\BBOQ}\APACrefatitle {{Causal mediation analysis for stochastic
  interventions}} {{Causal mediation analysis for stochastic
  interventions}}.{\BBCQ}
\newblock
\begin{APACrefURL} \url{https://arxiv.org/abs/1901.02776} \end{APACrefURL}
\PrintBackRefs{\CurrentBib}

\bibitem [\protect \citeauthoryear {%
Didelez%
}{%
Didelez%
}{%
{\protect \APACyear {2018}}%
}]{%
Didelez2018}
\APACinsertmetastar {%
Didelez2018}%
\begin{APACrefauthors}%
Didelez, V.%
\end{APACrefauthors}%
\unskip\
\newblock
\APACrefYearMonthDay{2018}{}{}.
\newblock
{\BBOQ}\APACrefatitle {{Defining causal meditation with a longitudinal mediator
  and a survival outcome}} {{Defining causal meditation with a longitudinal
  mediator and a survival outcome}}.{\BBCQ}
\newblock
\APACjournalVolNumPages{Lifetime Data Analysis}{}{}{}.
\newblock
\begin{APACrefDOI} \doi{10.1007/s10985-018-9449-0} \end{APACrefDOI}
\PrintBackRefs{\CurrentBib}

\bibitem [\protect \citeauthoryear {%
Didelez%
, Dawid%
\BCBL {}\ \BBA {} Geneletti%
}{%
Didelez%
\ \protect \BOthers {.}}{%
{\protect \APACyear {2006}}%
}]{%
Didelez2006}
\APACinsertmetastar {%
Didelez2006}%
\begin{APACrefauthors}%
Didelez, V.%
, Dawid, A\BPBI P.%
\BCBL {}\ \BBA {} Geneletti, S.%
\end{APACrefauthors}%
\unskip\
\newblock
\APACrefYearMonthDay{2006}{}{}.
\newblock
{\BBOQ}\APACrefatitle {Direct and Indirect Effects of Sequential Treatments}
  {Direct and indirect effects of sequential treatments}.{\BBCQ}
\newblock
\BIn{} \APACrefbtitle {{Proceedings of the 22nd Conference on Uncertainty in
  Artificial Intelligence}} {{Proceedings of the 22nd Conference on Uncertainty
  in Artificial Intelligence}}\ (\BPGS\ 138--146).
\newblock
\APACaddressPublisher{}{AUAI Press}.
\PrintBackRefs{\CurrentBib}

\bibitem [\protect \citeauthoryear {%
Ding%
\ \BBA {} Vanderweele%
}{%
Ding%
\ \BBA {} Vanderweele%
}{%
{\protect \APACyear {2016}}%
}]{%
Ding2016}
\APACinsertmetastar {%
Ding2016}%
\begin{APACrefauthors}%
Ding, P.%
\BCBT {}\ \BBA {} Vanderweele, T\BPBI J.%
\end{APACrefauthors}%
\unskip\
\newblock
\APACrefYearMonthDay{2016}{}{}.
\newblock
{\BBOQ}\APACrefatitle {{Sharp sensitivity bounds for mediation under unmeasured
  mediator-outcome confounding}} {{Sharp sensitivity bounds for mediation under
  unmeasured mediator-outcome confounding}}.{\BBCQ}
\newblock
\APACjournalVolNumPages{Biometrika}{103}{2}{483--490}.
\newblock
\begin{APACrefDOI} \doi{10.1093/biomet/asw012} \end{APACrefDOI}
\PrintBackRefs{\CurrentBib}

\bibitem [\protect \citeauthoryear {%
Frangakis%
\ \BBA {} Rubin%
}{%
Frangakis%
\ \BBA {} Rubin%
}{%
{\protect \APACyear {2002}}%
}]{%
Frangakis2002}
\APACinsertmetastar {%
Frangakis2002}%
\begin{APACrefauthors}%
Frangakis, C\BPBI E.%
\BCBT {}\ \BBA {} Rubin, D\BPBI B.%
\end{APACrefauthors}%
\unskip\
\newblock
\APACrefYearMonthDay{2002}{}{}.
\newblock
{\BBOQ}\APACrefatitle {{Principal Stratification in Causal Inference}}
  {{Principal Stratification in Causal Inference}}.{\BBCQ}
\newblock
\APACjournalVolNumPages{Biometrics}{58}{1}{21--29}.
\PrintBackRefs{\CurrentBib}

\bibitem [\protect \citeauthoryear {%
Geneletti%
}{%
Geneletti%
}{%
{\protect \APACyear {2007}}%
}]{%
Geneletti2007}
\APACinsertmetastar {%
Geneletti2007}%
\begin{APACrefauthors}%
Geneletti, S.%
\end{APACrefauthors}%
\unskip\
\newblock
\APACrefYearMonthDay{2007}{}{}.
\newblock
{\BBOQ}\APACrefatitle {{Identifying direct and indirect effects in a
  non-counterfactual framework}} {{Identifying direct and indirect effects in a
  non-counterfactual framework}}.{\BBCQ}
\newblock
\APACjournalVolNumPages{Journal of the Royal Statistical Society. Series B:
  Statistical Methodology}{69}{2}{199--215}.
\newblock
\begin{APACrefDOI} \doi{10.1111/j.1467-9868.2007.00584.x} \end{APACrefDOI}
\PrintBackRefs{\CurrentBib}

\bibitem [\protect \citeauthoryear {%
Glymour%
\ \BBA {} Glymour%
}{%
Glymour%
\ \BBA {} Glymour%
}{%
{\protect \APACyear {2014}}%
}]{%
Glymour2014}
\APACinsertmetastar {%
Glymour2014}%
\begin{APACrefauthors}%
Glymour, C.%
\BCBT {}\ \BBA {} Glymour, M\BPBI R.%
\end{APACrefauthors}%
\unskip\
\newblock
\APACrefYearMonthDay{2014}{}{}.
\newblock
{\BBOQ}\APACrefatitle {{Race and sex are causes}} {{Race and sex are
  causes}}.{\BBCQ}
\newblock
\APACjournalVolNumPages{Epidemiology}{25}{4}{488--490}.
\newblock
\begin{APACrefDOI} \doi{10.1097/EDE.0000000000000122} \end{APACrefDOI}
\PrintBackRefs{\CurrentBib}

\bibitem [\protect \citeauthoryear {%
Hayes%
}{%
Hayes%
}{%
{\protect \APACyear {2018}}%
}]{%
Hayes2018}
\APACinsertmetastar {%
Hayes2018}%
\begin{APACrefauthors}%
Hayes, A\BPBI F.%
\end{APACrefauthors}%
\unskip\
\newblock
\APACrefYear{2018}.
\newblock
\APACrefbtitle {{Introduction to Mediation, Moderation, and Conditional Process
  Analysis: A Regression-Based Approach}} {{Introduction to Mediation,
  Moderation, and Conditional Process Analysis: A Regression-Based Approach}}.
\newblock
\APACaddressPublisher{New York, NY}{The Guilford Press}.
\PrintBackRefs{\CurrentBib}

\bibitem [\protect \citeauthoryear {%
Holland%
}{%
Holland%
}{%
{\protect \APACyear {1986}}%
}]{%
Holland1986}
\APACinsertmetastar {%
Holland1986}%
\begin{APACrefauthors}%
Holland, P\BPBI W.%
\end{APACrefauthors}%
\unskip\
\newblock
\APACrefYearMonthDay{1986}{}{}.
\newblock
{\BBOQ}\APACrefatitle {Statistics and Causal Inference} {Statistics and causal
  inference}.{\BBCQ}
\newblock
\APACjournalVolNumPages{Journal of the American Statistical
  Association}{81}{396}{945}.
\newblock
\begin{APACrefDOI} \doi{10.2307/2289064} \end{APACrefDOI}
\PrintBackRefs{\CurrentBib}

\bibitem [\protect \citeauthoryear {%
Hong%
}{%
Hong%
}{%
{\protect \APACyear {2010}}%
}]{%
Hong2010}
\APACinsertmetastar {%
Hong2010}%
\begin{APACrefauthors}%
Hong, G.%
\end{APACrefauthors}%
\unskip\
\newblock
\APACrefYearMonthDay{2010}{}{}.
\newblock
{\BBOQ}\APACrefatitle {{Ratio of mediator probability weighting for estimating
  natural direct and indirect effects}} {{Ratio of mediator probability
  weighting for estimating natural direct and indirect effects}}.{\BBCQ}
\newblock
\BIn{} \APACrefbtitle {{Proceedings of the American Statistical Association,
  Biometrics Section}} {{Proceedings of the American Statistical Association,
  Biometrics Section}}\ (\BPGS\ 2401--2415).
\PrintBackRefs{\CurrentBib}

\bibitem [\protect \citeauthoryear {%
Hong%
, Deutsch%
\BCBL {}\ \BBA {} Hill%
}{%
Hong%
\ \protect \BOthers {.}}{%
{\protect \APACyear {2015}}%
}]{%
Hong2015}
\APACinsertmetastar {%
Hong2015}%
\begin{APACrefauthors}%
Hong, G.%
, Deutsch, J.%
\BCBL {}\ \BBA {} Hill, H\BPBI D.%
\end{APACrefauthors}%
\unskip\
\newblock
\APACrefYearMonthDay{2015}{}{}.
\newblock
{\BBOQ}\APACrefatitle {{Ratio-of-mediator-probability weighting for causal
  mediation analysis in the presence of treatment-by-mediator interaction}}
  {{Ratio-of-mediator-probability weighting for causal mediation analysis in
  the presence of treatment-by-mediator interaction}}.{\BBCQ}
\newblock
\APACjournalVolNumPages{Journal of Educational and Behavioral
  Statistics}{40}{3}{307--340}.
\newblock
\begin{APACrefDOI} \doi{10.3102/1076998615583902} \end{APACrefDOI}
\PrintBackRefs{\CurrentBib}

\bibitem [\protect \citeauthoryear {%
Imai%
, Jo%
\BCBL {}\ \BBA {} Stuart%
}{%
Imai%
\ \protect \BOthers {.}}{%
{\protect \APACyear {2011}}%
}]{%
Imai2011b}
\APACinsertmetastar {%
Imai2011b}%
\begin{APACrefauthors}%
Imai, K.%
, Jo, B.%
\BCBL {}\ \BBA {} Stuart, E\BPBI A.%
\end{APACrefauthors}%
\unskip\
\newblock
\APACrefYearMonthDay{2011}{}{}.
\newblock
{\BBOQ}\APACrefatitle {{Commentary: Using potential outcomes to understand
  causal mediation analysis}} {{Commentary: Using potential outcomes to
  understand causal mediation analysis}}.{\BBCQ}
\newblock
\APACjournalVolNumPages{Multivariate Behavioral Research}{46}{5}{861--873}.
\newblock
\begin{APACrefDOI} \doi{10.1080/00273171.2011.606743} \end{APACrefDOI}
\PrintBackRefs{\CurrentBib}

\bibitem [\protect \citeauthoryear {%
Imai%
, Keele%
\BCBL {}\ \BBA {} Tingley%
}{%
Imai%
, Keele%
\BCBL {}\ \BBA {} Tingley%
}{%
{\protect \APACyear {2010}}%
}]{%
Imai2010a}
\APACinsertmetastar {%
Imai2010a}%
\begin{APACrefauthors}%
Imai, K.%
, Keele, L.%
\BCBL {}\ \BBA {} Tingley, D.%
\end{APACrefauthors}%
\unskip\
\newblock
\APACrefYearMonthDay{2010}{}{}.
\newblock
{\BBOQ}\APACrefatitle {{A general approach to causal mediation analysis.}} {{A
  general approach to causal mediation analysis.}}{\BBCQ}
\newblock
\APACjournalVolNumPages{Psychological Methods}{15}{4}{309--34}.
\newblock
\begin{APACrefDOI} \doi{10.1037/a0020761} \end{APACrefDOI}
\PrintBackRefs{\CurrentBib}

\bibitem [\protect \citeauthoryear {%
Imai%
, Keele%
\BCBL {}\ \BBA {} Yamamoto%
}{%
Imai%
, Keele%
\BCBL {}\ \BBA {} Yamamoto%
}{%
{\protect \APACyear {2010}}%
}]{%
Imai2010}
\APACinsertmetastar {%
Imai2010}%
\begin{APACrefauthors}%
Imai, K.%
, Keele, L.%
\BCBL {}\ \BBA {} Yamamoto, T.%
\end{APACrefauthors}%
\unskip\
\newblock
\APACrefYearMonthDay{2010}{}{}.
\newblock
{\BBOQ}\APACrefatitle {{Identification, inference and sensitivity analysis for
  causal mediation effects}} {{Identification, inference and sensitivity
  analysis for causal mediation effects}}.{\BBCQ}
\newblock
\APACjournalVolNumPages{Statistical Science}{25}{1}{51--71}.
\newblock
\begin{APACrefDOI} \doi{10.1214/10-STS321} \end{APACrefDOI}
\PrintBackRefs{\CurrentBib}

\bibitem [\protect \citeauthoryear {%
Jackson%
}{%
Jackson%
}{%
{\protect \APACyear {2019}}%
}]{%
Jackson2019}
\APACinsertmetastar {%
Jackson2019}%
\begin{APACrefauthors}%
Jackson, J\BPBI W.%
\end{APACrefauthors}%
\unskip\
\newblock
\APACrefYearMonthDay{2019}{}{}.
\newblock
{\BBOQ}\APACrefatitle {{Meaningful causal decompositions in health equity
  research: definition, identification, and estimation through a weighting
  framework}} {{Meaningful causal decompositions in health equity research:
  definition, identification, and estimation through a weighting
  framework}}.{\BBCQ}
\newblock
\begin{APACrefURL} \url{http://arxiv.org/abs/1909.10060} \end{APACrefURL}
\PrintBackRefs{\CurrentBib}

\bibitem [\protect \citeauthoryear {%
Jackson%
\ \BBA {} VanderWeele%
}{%
Jackson%
\ \BBA {} VanderWeele%
}{%
{\protect \APACyear {2018}}%
}]{%
Jackson2018}
\APACinsertmetastar {%
Jackson2018}%
\begin{APACrefauthors}%
Jackson, J\BPBI W.%
\BCBT {}\ \BBA {} VanderWeele, T\BPBI J.%
\end{APACrefauthors}%
\unskip\
\newblock
\APACrefYearMonthDay{2018}{}{}.
\newblock
{\BBOQ}\APACrefatitle {Decomposition Analysis to Identify Intervention Targets
  for Reducing Disparities} {Decomposition analysis to identify intervention
  targets for reducing disparities}.{\BBCQ}
\newblock
\APACjournalVolNumPages{Epidemiology}{29}{6}{825--835}.
\newblock
\begin{APACrefDOI} \doi{10.1097/EDE.0000000000000901} \end{APACrefDOI}
\PrintBackRefs{\CurrentBib}

\bibitem [\protect \citeauthoryear {%
Jo%
\ \BBA {} Stuart%
}{%
Jo%
\ \BBA {} Stuart%
}{%
{\protect \APACyear {2009}}%
}]{%
Jo2009}
\APACinsertmetastar {%
Jo2009}%
\begin{APACrefauthors}%
Jo, B.%
\BCBT {}\ \BBA {} Stuart, E\BPBI A.%
\end{APACrefauthors}%
\unskip\
\newblock
\APACrefYearMonthDay{2009}{oct}{}.
\newblock
{\BBOQ}\APACrefatitle {{On the use of propensity scores in principal causal
  effect estimation}} {{On the use of propensity scores in principal causal
  effect estimation}}.{\BBCQ}
\newblock
\APACjournalVolNumPages{Statistics in Medicine}{28}{23}{2857--2875}.
\newblock
\begin{APACrefDOI} \doi{10.1002/sim.3669} \end{APACrefDOI}
\PrintBackRefs{\CurrentBib}

\bibitem [\protect \citeauthoryear {%
Jo%
\ \BBA {} Stuart%
}{%
Jo%
\ \BBA {} Stuart%
}{%
{\protect \APACyear {2012}}%
}]{%
Jo2012}
\APACinsertmetastar {%
Jo2012}%
\begin{APACrefauthors}%
Jo, B.%
\BCBT {}\ \BBA {} Stuart, E\BPBI A.%
\end{APACrefauthors}%
\unskip\
\newblock
\APACrefYearMonthDay{2012}{}{}.
\newblock
{\BBOQ}\APACrefatitle {{Comments: Causal Interpretations of Mediation Effects}}
  {{Comments: Causal Interpretations of Mediation Effects}}.{\BBCQ}
\newblock
\APACjournalVolNumPages{Journal of Research on Educational
  Effectiveness}{5}{3}{250--253}.
\newblock
\begin{APACrefDOI} \doi{10.1080/19345747.2012.688414} \end{APACrefDOI}
\PrintBackRefs{\CurrentBib}

\bibitem [\protect \citeauthoryear {%
Jo%
, Stuart%
, Mackinnon%
\BCBL {}\ \BBA {} Vinokur%
}{%
Jo%
\ \protect \BOthers {.}}{%
{\protect \APACyear {2011}}%
}]{%
Jo2011}
\APACinsertmetastar {%
Jo2011}%
\begin{APACrefauthors}%
Jo, B.%
, Stuart, E\BPBI A.%
, Mackinnon, D\BPBI P.%
\BCBL {}\ \BBA {} Vinokur, A\BPBI D.%
\end{APACrefauthors}%
\unskip\
\newblock
\APACrefYearMonthDay{2011}{}{}.
\newblock
{\BBOQ}\APACrefatitle {{The use of propensity scores in mediation analysis}}
  {{The use of propensity scores in mediation analysis}}.{\BBCQ}
\newblock
\APACjournalVolNumPages{Multivariate Behavioral Research}{46}{3}{425--452}.
\newblock
\begin{APACrefDOI} \doi{10.1080/00273171.2011.576624} \end{APACrefDOI}
\PrintBackRefs{\CurrentBib}

\bibitem [\protect \citeauthoryear {%
Judd%
\ \BBA {} Kenny%
}{%
Judd%
\ \BBA {} Kenny%
}{%
{\protect \APACyear {1981}}%
}]{%
Judd1981}
\APACinsertmetastar {%
Judd1981}%
\begin{APACrefauthors}%
Judd, C\BPBI M.%
\BCBT {}\ \BBA {} Kenny, D\BPBI A.%
\end{APACrefauthors}%
\unskip\
\newblock
\APACrefYearMonthDay{1981}{}{}.
\newblock
{\BBOQ}\APACrefatitle {{Process analysis: Estimating mediation in treatment
  evaluations}} {{Process analysis: Estimating mediation in treatment
  evaluations}}.{\BBCQ}
\newblock
\APACjournalVolNumPages{Evaluation Review}{5}{5}{602--619}.
\newblock
\begin{APACrefDOI} \doi{10.1017/CBO9781107415324.004} \end{APACrefDOI}
\PrintBackRefs{\CurrentBib}

\bibitem [\protect \citeauthoryear {%
Kennedy%
}{%
Kennedy%
}{%
{\protect \APACyear {2018}}%
}]{%
Kennedy2018}
\APACinsertmetastar {%
Kennedy2018}%
\begin{APACrefauthors}%
Kennedy, E\BPBI H.%
\end{APACrefauthors}%
\unskip\
\newblock
\APACrefYearMonthDay{2018}{}{}.
\newblock
{\BBOQ}\APACrefatitle {{Nonparametric Causal Effects Based on Incremental
  Propensity Score Interventions}} {{Nonparametric Causal Effects Based on
  Incremental Propensity Score Interventions}}.{\BBCQ}
\newblock
\APACjournalVolNumPages{Journal of the American Statistical Association}{}{}{}.
\newblock
\begin{APACrefDOI} \doi{10.1080/01621459.2017.1422737} \end{APACrefDOI}
\PrintBackRefs{\CurrentBib}

\bibitem [\protect \citeauthoryear {%
Lange%
\ \BBA {} Hansen%
}{%
Lange%
\ \BBA {} Hansen%
}{%
{\protect \APACyear {2011}}%
}]{%
Lange2011}
\APACinsertmetastar {%
Lange2011}%
\begin{APACrefauthors}%
Lange, T.%
\BCBT {}\ \BBA {} Hansen, J\BPBI V.%
\end{APACrefauthors}%
\unskip\
\newblock
\APACrefYearMonthDay{2011}{}{}.
\newblock
{\BBOQ}\APACrefatitle {{Direct and indirect effects in a survival context.}}
  {{Direct and indirect effects in a survival context.}}{\BBCQ}
\newblock
\APACjournalVolNumPages{Epidemiology}{22}{4}{575--581}.
\newblock
\begin{APACrefDOI} \doi{10.1097/EDE.0b013e31821c680c} \end{APACrefDOI}
\PrintBackRefs{\CurrentBib}

\bibitem [\protect \citeauthoryear {%
Lange%
, Vansteelandt%
\BCBL {}\ \BBA {} Bekaert%
}{%
Lange%
\ \protect \BOthers {.}}{%
{\protect \APACyear {2012}}%
}]{%
Lange2012}
\APACinsertmetastar {%
Lange2012}%
\begin{APACrefauthors}%
Lange, T.%
, Vansteelandt, S.%
\BCBL {}\ \BBA {} Bekaert, M.%
\end{APACrefauthors}%
\unskip\
\newblock
\APACrefYearMonthDay{2012}{}{}.
\newblock
{\BBOQ}\APACrefatitle {{A simple unified approach for estimating natural direct
  and indirect effects}} {{A simple unified approach for estimating natural
  direct and indirect effects}}.{\BBCQ}
\newblock
\APACjournalVolNumPages{American Journal of Epidemiology}{176}{3}{190--195}.
\newblock
\begin{APACrefDOI} \doi{10.1093/aje/kwr525} \end{APACrefDOI}
\PrintBackRefs{\CurrentBib}

\bibitem [\protect \citeauthoryear {%
Lok%
}{%
Lok%
}{%
{\protect \APACyear {2016}}%
}]{%
Lok2016}
\APACinsertmetastar {%
Lok2016}%
\begin{APACrefauthors}%
Lok, J\BPBI J.%
\end{APACrefauthors}%
\unskip\
\newblock
\APACrefYearMonthDay{2016}{}{}.
\newblock
{\BBOQ}\APACrefatitle {{Defining and estimating causal direct and indirect
  effects when setting the mediator to specific values is not feasible}}
  {{Defining and estimating causal direct and indirect effects when setting the
  mediator to specific values is not feasible}}.{\BBCQ}
\newblock
\APACjournalVolNumPages{Statistics in Medicine}{35}{22}{4008--4020}.
\newblock
\begin{APACrefDOI} \doi{10.1002/sim.6990} \end{APACrefDOI}
\PrintBackRefs{\CurrentBib}

\bibitem [\protect \citeauthoryear {%
MacKinnon%
}{%
MacKinnon%
}{%
{\protect \APACyear {2008}}%
}]{%
MacKinnon2008}
\APACinsertmetastar {%
MacKinnon2008}%
\begin{APACrefauthors}%
MacKinnon, D\BPBI P.%
\end{APACrefauthors}%
\unskip\
\newblock
\APACrefYear{2008}.
\newblock
\APACrefbtitle {{Introduction to Statistical Mediation Analysis}}
  {{Introduction to Statistical Mediation Analysis}}.
\newblock
\APACaddressPublisher{New York, NY}{Taylor {\&} Francis}.
\PrintBackRefs{\CurrentBib}

\bibitem [\protect \citeauthoryear {%
Mackinnon%
, Fairchild%
\BCBL {}\ \BBA {} Fritz%
}{%
Mackinnon%
\ \protect \BOthers {.}}{%
{\protect \APACyear {2007}}%
}]{%
MacKinnon2007}
\APACinsertmetastar {%
MacKinnon2007}%
\begin{APACrefauthors}%
Mackinnon, D\BPBI P.%
, Fairchild, A\BPBI J.%
\BCBL {}\ \BBA {} Fritz, M\BPBI S.%
\end{APACrefauthors}%
\unskip\
\newblock
\APACrefYearMonthDay{2007}{}{}.
\newblock
{\BBOQ}\APACrefatitle {Mediation Analysis} {Mediation analysis}.{\BBCQ}
\newblock
\APACjournalVolNumPages{Annual Review of Psychology}{58}{}{593--614}.
\newblock
\begin{APACrefDOI} \doi{10.1146/annurev.psych.58.110405.085542}
  \end{APACrefDOI}
\PrintBackRefs{\CurrentBib}

\bibitem [\protect \citeauthoryear {%
MacKinnon%
, Kisbu-Sakarya%
\BCBL {}\ \BBA {} Gottschall%
}{%
MacKinnon%
\ \protect \BOthers {.}}{%
{\protect \APACyear {2013}}%
}]{%
MacKinnon2013}
\APACinsertmetastar {%
MacKinnon2013}%
\begin{APACrefauthors}%
MacKinnon, D\BPBI P.%
, Kisbu-Sakarya, Y.%
\BCBL {}\ \BBA {} Gottschall, A\BPBI C.%
\end{APACrefauthors}%
\unskip\
\newblock
\APACrefYearMonthDay{2013}{}{}.
\newblock
{\BBOQ}\APACrefatitle {Developments in Mediation Analysis} {Developments in
  mediation analysis}.{\BBCQ}
\newblock
\BIn{} \APACrefbtitle {{The Oxford Handbook of Quantitative Methods in
  Psychology: Vol. 2: Statistical Analysis}.} {{The Oxford Handbook of
  Quantitative Methods in Psychology: Vol. 2: Statistical Analysis}.}
\newblock
\begin{APACrefDOI} \doi{10.1093/oxfordhb/9780199934898.013.0016}
  \end{APACrefDOI}
\PrintBackRefs{\CurrentBib}

\bibitem [\protect \citeauthoryear {%
MacKinnon%
\ \BBA {} Pirlott%
}{%
MacKinnon%
\ \BBA {} Pirlott%
}{%
{\protect \APACyear {2015}}%
}]{%
MacKinnon2015}
\APACinsertmetastar {%
MacKinnon2015}%
\begin{APACrefauthors}%
MacKinnon, D\BPBI P.%
\BCBT {}\ \BBA {} Pirlott, A\BPBI G.%
\end{APACrefauthors}%
\unskip\
\newblock
\APACrefYearMonthDay{2015}{}{}.
\newblock
{\BBOQ}\APACrefatitle {Statistical Approaches for Enhancing Causal
  Interpretation of the {M} to {Y} Relation in Mediation Analysis} {Statistical
  approaches for enhancing causal interpretation of the {M} to {Y} relation in
  mediation analysis}.{\BBCQ}
\newblock
\APACjournalVolNumPages{Personality and Social Psychology
  Review}{19}{1}{30--43}.
\newblock
\begin{APACrefDOI} \doi{10.1177/1088868314542878} \end{APACrefDOI}
\PrintBackRefs{\CurrentBib}

\bibitem [\protect \citeauthoryear {%
MacKinnon%
, Valente%
\BCBL {}\ \BBA {} Gonzalez%
}{%
MacKinnon%
\ \protect \BOthers {.}}{%
{\protect \APACyear {2019}}%
}]{%
MacKinnon2019}
\APACinsertmetastar {%
MacKinnon2019}%
\begin{APACrefauthors}%
MacKinnon, D\BPBI P.%
, Valente, M\BPBI J.%
\BCBL {}\ \BBA {} Gonzalez, O.%
\end{APACrefauthors}%
\unskip\
\newblock
\APACrefYearMonthDay{2019}{}{}.
\newblock
{\BBOQ}\APACrefatitle {{The Correspondence Between Causal and Traditional
  Mediation Analysis: the Link Is the Mediator by Treatment Interaction}} {{The
  Correspondence Between Causal and Traditional Mediation Analysis: the Link Is
  the Mediator by Treatment Interaction}}.{\BBCQ}
\newblock
\APACjournalVolNumPages{Prevention Science}{}{}{}.
\newblock
\begin{APACrefDOI} \doi{10.1007/s11121-019-01076-4} \end{APACrefDOI}
\PrintBackRefs{\CurrentBib}

\bibitem [\protect \citeauthoryear {%
Mio{\v{c}}evi{\'{c}}%
, Gonzalez%
, Valente%
\BCBL {}\ \BBA {} MacKinnon%
}{%
Mio{\v{c}}evi{\'{c}}%
\ \protect \BOthers {.}}{%
{\protect \APACyear {2018}}%
}]{%
Miocevic2018}
\APACinsertmetastar {%
Miocevic2018}%
\begin{APACrefauthors}%
Mio{\v{c}}evi{\'{c}}, M.%
, Gonzalez, O.%
, Valente, M\BPBI J.%
\BCBL {}\ \BBA {} MacKinnon, D\BPBI P.%
\end{APACrefauthors}%
\unskip\
\newblock
\APACrefYearMonthDay{2018}{}{}.
\newblock
{\BBOQ}\APACrefatitle {A Tutorial in {Bayesian} Potential Outcomes Mediation
  Analysis} {A tutorial in {Bayesian} potential outcomes mediation
  analysis}.{\BBCQ}
\newblock
\APACjournalVolNumPages{Structural Equation Modeling}{25}{1}{121--136}.
\newblock
\begin{APACrefDOI} \doi{10.1080/10705511.2017.1342541} \end{APACrefDOI}
\PrintBackRefs{\CurrentBib}

\bibitem [\protect \citeauthoryear {%
Muth{\'{e}}n%
}{%
Muth{\'{e}}n%
}{%
{\protect \APACyear {2011}}%
}]{%
Muthen2011}
\APACinsertmetastar {%
Muthen2011}%
\begin{APACrefauthors}%
Muth{\'{e}}n, B\BPBI O.%
\end{APACrefauthors}%
\unskip\
\newblock
\APACrefYearMonthDay{2011}{}{}.
\newblock
{\BBOQ}\APACrefatitle {Applications of Causally Defined Direct and Indirect
  Effects in Mediation Analysis using {SEM} in {Mplus}} {Applications of
  causally defined direct and indirect effects in mediation analysis using
  {SEM} in {Mplus}}.{\BBCQ}
\newblock
\begin{APACrefURL} \url{http://www.statmodel2.com/download/causalmediation.pdf}
  \end{APACrefURL}
\PrintBackRefs{\CurrentBib}

\bibitem [\protect \citeauthoryear {%
Muth{\'{e}}n%
\ \BBA {} Asparouhov%
}{%
Muth{\'{e}}n%
\ \BBA {} Asparouhov%
}{%
{\protect \APACyear {2015}}%
}]{%
Muthen2015a}
\APACinsertmetastar {%
Muthen2015a}%
\begin{APACrefauthors}%
Muth{\'{e}}n, B\BPBI O.%
\BCBT {}\ \BBA {} Asparouhov, T.%
\end{APACrefauthors}%
\unskip\
\newblock
\APACrefYearMonthDay{2015}{}{}.
\newblock
{\BBOQ}\APACrefatitle {{Causal effects in mediation modeling: An introduction
  with applications to latent variables}} {{Causal effects in mediation
  modeling: An introduction with applications to latent variables}}.{\BBCQ}
\newblock
\APACjournalVolNumPages{Structural Equation Modeling}{22}{1}{12--23}.
\newblock
\begin{APACrefDOI} \doi{10.1080/10705511.2014.935843} \end{APACrefDOI}
\PrintBackRefs{\CurrentBib}

\bibitem [\protect \citeauthoryear {%
{NIH}%
}{%
{NIH}%
}{%
{\protect \APACyear {2016}}%
}]{%
RFA}
\APACinsertmetastar {%
RFA}%
\begin{APACrefauthors}%
{NIH}.%
\end{APACrefauthors}%
\unskip\
\newblock
\APACrefYearMonthDay{2016}{}{}.
\newblock
\APACrefbtitle {{RFA-MH-17-608}: Clinical Trials to Test the Effectiveness of
  Treatment, Preventive, and Services Interventions ({R01}).} {{RFA-MH-17-608}:
  Clinical trials to test the effectiveness of treatment, preventive, and
  services interventions ({R01}).}
\newblock
\APAChowpublished
  {\url{https://grants.nih.gov/grants/guide/rfa-files/rfa-mh-17-608.html}}.
\PrintBackRefs{\CurrentBib}

\bibitem [\protect \citeauthoryear {%
Pearl%
}{%
Pearl%
}{%
{\protect \APACyear {2001}}%
}]{%
Pearl2001}
\APACinsertmetastar {%
Pearl2001}%
\begin{APACrefauthors}%
Pearl, J.%
\end{APACrefauthors}%
\unskip\
\newblock
\APACrefYearMonthDay{2001}{}{}.
\newblock
{\BBOQ}\APACrefatitle {{Direct and indirect effects}} {{Direct and indirect
  effects}}.{\BBCQ}
\newblock
\APACjournalVolNumPages{Proceedings of the Seventeenth Conference on
  Uncertainty and Artificial Intelligence}{}{}{411--420}.
\PrintBackRefs{\CurrentBib}

\bibitem [\protect \citeauthoryear {%
Pearl%
}{%
Pearl%
}{%
{\protect \APACyear {2009}}%
}]{%
Pearl2009}
\APACinsertmetastar {%
Pearl2009}%
\begin{APACrefauthors}%
Pearl, J.%
\end{APACrefauthors}%
\unskip\
\newblock
\APACrefYearMonthDay{2009}{}{}.
\newblock
{\BBOQ}\APACrefatitle {{Causal inference in statistics: An overview}} {{Causal
  inference in statistics: An overview}}.{\BBCQ}
\newblock
\APACjournalVolNumPages{Statistics Surveys}{3}{}{96--146}.
\newblock
\begin{APACrefDOI} \doi{10.1214/09-SS057} \end{APACrefDOI}
\PrintBackRefs{\CurrentBib}

\bibitem [\protect \citeauthoryear {%
Pearl%
}{%
Pearl%
}{%
{\protect \APACyear {2012}}%
}]{%
Pearl2012}
\APACinsertmetastar {%
Pearl2012}%
\begin{APACrefauthors}%
Pearl, J.%
\end{APACrefauthors}%
\unskip\
\newblock
\APACrefYearMonthDay{2012}{}{}.
\newblock
{\BBOQ}\APACrefatitle {{The causal mediation formula--a guide to the assessment
  of pathways and mechanisms.}} {{The causal mediation formula--a guide to the
  assessment of pathways and mechanisms.}}{\BBCQ}
\newblock
\APACjournalVolNumPages{Prevention Science}{13}{4}{426--36}.
\newblock
\begin{APACrefDOI} \doi{10.1007/s11121-011-0270-1} \end{APACrefDOI}
\PrintBackRefs{\CurrentBib}

\bibitem [\protect \citeauthoryear {%
Petersen%
, Sinisi%
\BCBL {}\ \BBA {} van~der Laan%
}{%
Petersen%
\ \protect \BOthers {.}}{%
{\protect \APACyear {2006}}%
}]{%
Petersen2006}
\APACinsertmetastar {%
Petersen2006}%
\begin{APACrefauthors}%
Petersen, M\BPBI L.%
, Sinisi, S\BPBI E.%
\BCBL {}\ \BBA {} van~der Laan, M\BPBI J.%
\end{APACrefauthors}%
\unskip\
\newblock
\APACrefYearMonthDay{2006}{}{}.
\newblock
{\BBOQ}\APACrefatitle {{Estimation of direct causal effects.}} {{Estimation of
  direct causal effects.}}{\BBCQ}
\newblock
\APACjournalVolNumPages{Epidemiology}{17}{3}{276--84}.
\newblock
\begin{APACrefDOI} \doi{10.1097/01.ede.0000208475.99429.2d} \end{APACrefDOI}
\PrintBackRefs{\CurrentBib}

\bibitem [\protect \citeauthoryear {%
Preacher%
}{%
Preacher%
}{%
{\protect \APACyear {2015}}%
}]{%
Preacher2015}
\APACinsertmetastar {%
Preacher2015}%
\begin{APACrefauthors}%
Preacher, K\BPBI J.%
\end{APACrefauthors}%
\unskip\
\newblock
\APACrefYearMonthDay{2015}{}{}.
\newblock
{\BBOQ}\APACrefatitle {{Advances in Mediation Analysis: A Survey and Synthesis
  of New Developments}} {{Advances in Mediation Analysis: A Survey and
  Synthesis of New Developments}}.{\BBCQ}
\newblock
\APACjournalVolNumPages{Annual Review of Psychology}{66}{1}{825--852}.
\newblock
\begin{APACrefDOI} \doi{10.1146/annurev-psych-010814-015258} \end{APACrefDOI}
\PrintBackRefs{\CurrentBib}

\bibitem [\protect \citeauthoryear {%
Robins%
}{%
Robins%
}{%
{\protect \APACyear {2003}}%
}]{%
Robins2003}
\APACinsertmetastar {%
Robins2003}%
\begin{APACrefauthors}%
Robins, J\BPBI M.%
\end{APACrefauthors}%
\unskip\
\newblock
\APACrefYearMonthDay{2003}{}{}.
\newblock
{\BBOQ}\APACrefatitle {{Semantics of causal DAG models and the identification
  of direct and indirect effects}} {{Semantics of causal DAG models and the
  identification of direct and indirect effects}}.{\BBCQ}
\newblock
\BIn{} P.~Green, N.~Hjort\BCBL {}\ \BBA {} S.~Richardson\ (\BEDS),
  \APACrefbtitle {{Highly Structured Stochastic Systems}} {{Highly Structured
  Stochastic Systems}}\ (\BPGS\ 70--81).
\newblock
\APACaddressPublisher{Oxford, UK}{Oxford University Press}.
\PrintBackRefs{\CurrentBib}

\bibitem [\protect \citeauthoryear {%
Robins%
\ \BBA {} Greenland%
}{%
Robins%
\ \BBA {} Greenland%
}{%
{\protect \APACyear {1992}}%
}]{%
Robins1992}
\APACinsertmetastar {%
Robins1992}%
\begin{APACrefauthors}%
Robins, J\BPBI M.%
\BCBT {}\ \BBA {} Greenland, S.%
\end{APACrefauthors}%
\unskip\
\newblock
\APACrefYearMonthDay{1992}{}{}.
\newblock
{\BBOQ}\APACrefatitle {{Identifiability and exchangeability for direct and
  indirect effects}} {{Identifiability and exchangeability for direct and
  indirect effects}}.{\BBCQ}
\newblock
\APACjournalVolNumPages{Epidemiology}{3}{2}{143--155}.
\PrintBackRefs{\CurrentBib}

\bibitem [\protect \citeauthoryear {%
Rubin%
}{%
Rubin%
}{%
{\protect \APACyear {1974}}%
}]{%
Rubin1974}
\APACinsertmetastar {%
Rubin1974}%
\begin{APACrefauthors}%
Rubin, D\BPBI B.%
\end{APACrefauthors}%
\unskip\
\newblock
\APACrefYearMonthDay{1974}{}{}.
\newblock
{\BBOQ}\APACrefatitle {{Estimating causal effects of treatments in randomized
  and nonrandomized studies.}} {{Estimating causal effects of treatments in
  randomized and nonrandomized studies.}}{\BBCQ}
\newblock
\APACjournalVolNumPages{Journal of Educational Psychology}{66}{5}{688--701}.
\newblock
\begin{APACrefDOI} \doi{10.1037/h0037350} \end{APACrefDOI}
\PrintBackRefs{\CurrentBib}

\bibitem [\protect \citeauthoryear {%
Rudolph%
, Sofrygin%
, Zheng%
\BCBL {}\ \BBA {} van~der Laan%
}{%
Rudolph%
\ \protect \BOthers {.}}{%
{\protect \APACyear {2017}}%
}]{%
Rudolph2017}
\APACinsertmetastar {%
Rudolph2017}%
\begin{APACrefauthors}%
Rudolph, K\BPBI E.%
, Sofrygin, O.%
, Zheng, W.%
\BCBL {}\ \BBA {} van~der Laan, M\BPBI J.%
\end{APACrefauthors}%
\unskip\
\newblock
\APACrefYearMonthDay{2017}{}{}.
\newblock
{\BBOQ}\APACrefatitle {{Robust and flexible estimation of data-dependent
  stochastic mediation effects: A proposed method and example in a randomized
  trial setting}} {{Robust and flexible estimation of data-dependent stochastic
  mediation effects: A proposed method and example in a randomized trial
  setting}}.{\BBCQ}
\newblock
\APACjournalVolNumPages{Epidemiologic Methods}{7}{}{}.
\newblock
\begin{APACrefDOI} \doi{10.1515/em-2017-0007} \end{APACrefDOI}
\PrintBackRefs{\CurrentBib}

\bibitem [\protect \citeauthoryear {%
Splawa-Neyman%
}{%
Splawa-Neyman%
}{%
{\protect \APACyear {1923}}%
}]{%
Splawa-Neyman1923}
\APACinsertmetastar {%
Splawa-Neyman1923}%
\begin{APACrefauthors}%
Splawa-Neyman, J.%
\end{APACrefauthors}%
\unskip\
\newblock
\APACrefYearMonthDay{1923}{}{}.
\newblock
{\BBOQ}\APACrefatitle {{On the application of probability theory to
  agricultural experiments. Essay on principles. Section 9 [Translated and
  edited by D. M. Dabrowska and T. P. Speed]}} {{On the application of
  probability theory to agricultural experiments. Essay on principles. Section
  9 [Translated and edited by D. M. Dabrowska and T. P. Speed]}}.{\BBCQ}
\newblock
\APACjournalVolNumPages{Annals of Agricultural Sciences}{10}{}{1--51}.
\newblock
\begin{APACrefDOI} \doi{10.1214/ss/1177012031} \end{APACrefDOI}
\PrintBackRefs{\CurrentBib}

\bibitem [\protect \citeauthoryear {%
{Tchetgen Tchetgen}%
}{%
{Tchetgen Tchetgen}%
}{%
{\protect \APACyear {2011}}%
}]{%
TchetgenTchetgen2011}
\APACinsertmetastar {%
TchetgenTchetgen2011}%
\begin{APACrefauthors}%
{Tchetgen Tchetgen}, E\BPBI J.%
\end{APACrefauthors}%
\unskip\
\newblock
\APACrefYearMonthDay{2011}{}{}.
\newblock
{\BBOQ}\APACrefatitle {{On causal mediation analysis with a survival outcome}}
  {{On causal mediation analysis with a survival outcome}}.{\BBCQ}
\newblock
\APACjournalVolNumPages{International Journal of Biostatistics}{7}{1}{}.
\newblock
\begin{APACrefDOI} \doi{10.2202/1557-4679.1351} \end{APACrefDOI}
\PrintBackRefs{\CurrentBib}

\bibitem [\protect \citeauthoryear {%
{Tchetgen Tchetgen}%
\ \BBA {} Shpitser%
}{%
{Tchetgen Tchetgen}%
\ \BBA {} Shpitser%
}{%
{\protect \APACyear {2012}}%
}]{%
TchetgenTchetgen2012}
\APACinsertmetastar {%
TchetgenTchetgen2012}%
\begin{APACrefauthors}%
{Tchetgen Tchetgen}, E\BPBI J.%
\BCBT {}\ \BBA {} Shpitser, I.%
\end{APACrefauthors}%
\unskip\
\newblock
\APACrefYearMonthDay{2012}{}{}.
\newblock
{\BBOQ}\APACrefatitle {{Semiparametric theory for causal mediation analysis:
  Efficiency bounds, multiple robustness and sensitivity analysis}}
  {{Semiparametric theory for causal mediation analysis: Efficiency bounds,
  multiple robustness and sensitivity analysis}}.{\BBCQ}
\newblock
\APACjournalVolNumPages{The Annals of Statistics}{40}{3}{1816--1845}.
\newblock
\begin{APACrefDOI} \doi{10.1214/12-AOS990} \end{APACrefDOI}
\PrintBackRefs{\CurrentBib}

\bibitem [\protect \citeauthoryear {%
Tingley%
, Yamamoto%
, Hirose%
, Keele%
\BCBL {}\ \BBA {} Imai%
}{%
Tingley%
\ \protect \BOthers {.}}{%
{\protect \APACyear {2014}}%
}]{%
Tingley2014}
\APACinsertmetastar {%
Tingley2014}%
\begin{APACrefauthors}%
Tingley, D.%
, Yamamoto, T.%
, Hirose, K.%
, Keele, L.%
\BCBL {}\ \BBA {} Imai, K.%
\end{APACrefauthors}%
\unskip\
\newblock
\APACrefYearMonthDay{2014}{}{}.
\newblock
{\BBOQ}\APACrefatitle {{mediation: R package for causal mediation analysis}}
  {{mediation: R package for causal mediation analysis}}.{\BBCQ}
\newblock
\APACjournalVolNumPages{Journal of Statistical Software}{59}{5}{1--38}.
\newblock
\begin{APACrefDOI} \doi{10.18637/jss.v059.i05} \end{APACrefDOI}
\PrintBackRefs{\CurrentBib}

\bibitem [\protect \citeauthoryear {%
Valeri%
\ \BBA {} VanderWeele%
}{%
Valeri%
\ \BBA {} VanderWeele%
}{%
{\protect \APACyear {2013}}%
}]{%
Valeri2013}
\APACinsertmetastar {%
Valeri2013}%
\begin{APACrefauthors}%
Valeri, L.%
\BCBT {}\ \BBA {} VanderWeele, T\BPBI J.%
\end{APACrefauthors}%
\unskip\
\newblock
\APACrefYearMonthDay{2013}{}{}.
\newblock
{\BBOQ}\APACrefatitle {{Mediation analysis allowing for exposure–mediator
  interactions and causal interpretation: Theoretical assumptions and
  implementation with SAS and SPSS macros.}} {{Mediation analysis allowing for
  exposure–mediator interactions and causal interpretation: Theoretical
  assumptions and implementation with SAS and SPSS macros.}}{\BBCQ}
\newblock
\APACjournalVolNumPages{Psychological methods}{18}{2}{137--150}.
\newblock
\begin{APACrefDOI} \doi{10.1037/a0031034} \end{APACrefDOI}
\PrintBackRefs{\CurrentBib}

\bibitem [\protect \citeauthoryear {%
Valeri%
\ \BBA {} VanderWeele%
}{%
Valeri%
\ \BBA {} VanderWeele%
}{%
{\protect \APACyear {2015}}%
}]{%
Valeri2015}
\APACinsertmetastar {%
Valeri2015}%
\begin{APACrefauthors}%
Valeri, L.%
\BCBT {}\ \BBA {} VanderWeele, T\BPBI J.%
\end{APACrefauthors}%
\unskip\
\newblock
\APACrefYearMonthDay{2015}{}{}.
\newblock
{\BBOQ}\APACrefatitle {{SAS macro for causal mediation analysis with survival
  data}} {{SAS macro for causal mediation analysis with survival data}}.{\BBCQ}
\newblock
\APACjournalVolNumPages{Epidemiology}{26}{2}{e23--e24}.
\newblock
\begin{APACrefDOI} \doi{10.1097/ede.0000000000000253} \end{APACrefDOI}
\PrintBackRefs{\CurrentBib}

\bibitem [\protect \citeauthoryear {%
VanderWeele%
}{%
VanderWeele%
}{%
{\protect \APACyear {2011}}%
}]{%
Vanderweele2011}
\APACinsertmetastar {%
Vanderweele2011}%
\begin{APACrefauthors}%
VanderWeele, T\BPBI J.%
\end{APACrefauthors}%
\unskip\
\newblock
\APACrefYearMonthDay{2011}{}{}.
\newblock
{\BBOQ}\APACrefatitle {{Causal mediation analysis with survival data}} {{Causal
  mediation analysis with survival data}}.{\BBCQ}
\newblock
\APACjournalVolNumPages{Epidemiology}{22}{4}{582--585}.
\newblock
\begin{APACrefDOI} \doi{10.1097/EDE.0b013e31821db37e} \end{APACrefDOI}
\PrintBackRefs{\CurrentBib}

\bibitem [\protect \citeauthoryear {%
VanderWeele%
\ \BBA {} {Tchetgen Tchetgen}%
}{%
VanderWeele%
\ \BBA {} {Tchetgen Tchetgen}%
}{%
{\protect \APACyear {2017}}%
}]{%
VanderWeele2017}
\APACinsertmetastar {%
VanderWeele2017}%
\begin{APACrefauthors}%
VanderWeele, T\BPBI J.%
\BCBT {}\ \BBA {} {Tchetgen Tchetgen}, E\BPBI J.%
\end{APACrefauthors}%
\unskip\
\newblock
\APACrefYearMonthDay{2017}{}{}.
\newblock
{\BBOQ}\APACrefatitle {{Mediation analysis with time varying exposures and
  mediators}} {{Mediation analysis with time varying exposures and
  mediators}}.{\BBCQ}
\newblock
\APACjournalVolNumPages{Journal of the Royal Statistical Society. Series B:
  Statistical Methodology}{79}{3}{917--938}.
\newblock
\begin{APACrefDOI} \doi{10.1111/rssb.12194} \end{APACrefDOI}
\PrintBackRefs{\CurrentBib}

\bibitem [\protect \citeauthoryear {%
VanderWeele%
\ \BBA {} Vansteelandt%
}{%
VanderWeele%
\ \BBA {} Vansteelandt%
}{%
{\protect \APACyear {2009}}%
}]{%
VanderWeele2009}
\APACinsertmetastar {%
VanderWeele2009}%
\begin{APACrefauthors}%
VanderWeele, T\BPBI J.%
\BCBT {}\ \BBA {} Vansteelandt, S.%
\end{APACrefauthors}%
\unskip\
\newblock
\APACrefYearMonthDay{2009}{}{}.
\newblock
{\BBOQ}\APACrefatitle {{Conceptual issues concerning mediation, interventions
  and composition}} {{Conceptual issues concerning mediation, interventions and
  composition}}.{\BBCQ}
\newblock
\APACjournalVolNumPages{Statistics and its Interface}{2}{}{457--468}.
\PrintBackRefs{\CurrentBib}

\bibitem [\protect \citeauthoryear {%
VanderWeele%
\ \BBA {} Vansteelandt%
}{%
VanderWeele%
\ \BBA {} Vansteelandt%
}{%
{\protect \APACyear {2010}}%
}]{%
VanderWeele2010}
\APACinsertmetastar {%
VanderWeele2010}%
\begin{APACrefauthors}%
VanderWeele, T\BPBI J.%
\BCBT {}\ \BBA {} Vansteelandt, S.%
\end{APACrefauthors}%
\unskip\
\newblock
\APACrefYearMonthDay{2010}{}{}.
\newblock
{\BBOQ}\APACrefatitle {{Odds ratios for mediation analysis for a dichotomous
  outcome}} {{Odds ratios for mediation analysis for a dichotomous
  outcome}}.{\BBCQ}
\newblock
\APACjournalVolNumPages{American Journal of Epidemiology}{172}{12}{1339--1348}.
\newblock
\begin{APACrefDOI} \doi{10.1093/aje/kwq332} \end{APACrefDOI}
\PrintBackRefs{\CurrentBib}

\bibitem [\protect \citeauthoryear {%
VanderWeele%
\ \BBA {} Vansteelandt%
}{%
VanderWeele%
\ \BBA {} Vansteelandt%
}{%
{\protect \APACyear {2013}}%
}]{%
VanderWeele2013a}
\APACinsertmetastar {%
VanderWeele2013a}%
\begin{APACrefauthors}%
VanderWeele, T\BPBI J.%
\BCBT {}\ \BBA {} Vansteelandt, S.%
\end{APACrefauthors}%
\unskip\
\newblock
\APACrefYearMonthDay{2013}{}{}.
\newblock
{\BBOQ}\APACrefatitle {{Mediation analysis with multiple mediators}}
  {{Mediation analysis with multiple mediators}}.{\BBCQ}
\newblock
\APACjournalVolNumPages{Epidemiologic Methods}{2}{1}{95--115}.
\newblock
\begin{APACrefDOI} \doi{10.1515/em-2012-0010} \end{APACrefDOI}
\PrintBackRefs{\CurrentBib}

\bibitem [\protect \citeauthoryear {%
VanderWeele%
, Vansteelandt%
\BCBL {}\ \BBA {} Robins%
}{%
VanderWeele%
\ \protect \BOthers {.}}{%
{\protect \APACyear {2014}}%
}]{%
VanderWeele2014a}
\APACinsertmetastar {%
VanderWeele2014a}%
\begin{APACrefauthors}%
VanderWeele, T\BPBI J.%
, Vansteelandt, S.%
\BCBL {}\ \BBA {} Robins, J\BPBI M.%
\end{APACrefauthors}%
\unskip\
\newblock
\APACrefYearMonthDay{2014}{}{}.
\newblock
{\BBOQ}\APACrefatitle {{Effect decomposition in the presence of an
  exposure-induced mediator-outcome confounder.}} {{Effect decomposition in the
  presence of an exposure-induced mediator-outcome confounder.}}{\BBCQ}
\newblock
\APACjournalVolNumPages{Epidemiology}{25}{2}{300--6}.
\newblock
\begin{APACrefDOI} \doi{10.1097/EDE.0000000000000034} \end{APACrefDOI}
\PrintBackRefs{\CurrentBib}

\bibitem [\protect \citeauthoryear {%
Vansteelandt%
, Bekaert%
\BCBL {}\ \BBA {} Lange%
}{%
Vansteelandt%
\ \protect \BOthers {.}}{%
{\protect \APACyear {2012}}%
}]{%
Vansteelandt2012a}
\APACinsertmetastar {%
Vansteelandt2012a}%
\begin{APACrefauthors}%
Vansteelandt, S.%
, Bekaert, M.%
\BCBL {}\ \BBA {} Lange, T.%
\end{APACrefauthors}%
\unskip\
\newblock
\APACrefYearMonthDay{2012}{}{}.
\newblock
{\BBOQ}\APACrefatitle {{Imputation strategies for the estimation of natural
  direct and indirect effects}} {{Imputation strategies for the estimation of
  natural direct and indirect effects}}.{\BBCQ}
\newblock
\APACjournalVolNumPages{Epidemiologic Methods}{1}{1}{7}.
\newblock
\begin{APACrefDOI} \doi{10.1515/2161-962X.1014} \end{APACrefDOI}
\PrintBackRefs{\CurrentBib}

\bibitem [\protect \citeauthoryear {%
Vansteelandt%
\ \BBA {} Daniel%
}{%
Vansteelandt%
\ \BBA {} Daniel%
}{%
{\protect \APACyear {2017}}%
}]{%
Vansteelandt2017}
\APACinsertmetastar {%
Vansteelandt2017}%
\begin{APACrefauthors}%
Vansteelandt, S.%
\BCBT {}\ \BBA {} Daniel, R\BPBI M.%
\end{APACrefauthors}%
\unskip\
\newblock
\APACrefYearMonthDay{2017}{}{}.
\newblock
{\BBOQ}\APACrefatitle {{Interventional effects for mediation analysis with
  multiple mediators}} {{Interventional effects for mediation analysis with
  multiple mediators}}.{\BBCQ}
\newblock
\APACjournalVolNumPages{Epidemiology}{28}{2}{258--265}.
\newblock
\begin{APACrefDOI} \doi{10.1097/EDE.0000000000000596} \end{APACrefDOI}
\PrintBackRefs{\CurrentBib}

\bibitem [\protect \citeauthoryear {%
Vansteelandt%
\ \BBA {} VanderWeele%
}{%
Vansteelandt%
\ \BBA {} VanderWeele%
}{%
{\protect \APACyear {2012}}%
}]{%
Vansteelandt2012}
\APACinsertmetastar {%
Vansteelandt2012}%
\begin{APACrefauthors}%
Vansteelandt, S.%
\BCBT {}\ \BBA {} VanderWeele, T\BPBI J.%
\end{APACrefauthors}%
\unskip\
\newblock
\APACrefYearMonthDay{2012}{}{}.
\newblock
{\BBOQ}\APACrefatitle {{Natural direct and indirect effects on the exposed:
  Effect decomposition under weaker assumptions}} {{Natural direct and indirect
  effects on the exposed: Effect decomposition under weaker
  assumptions}}.{\BBCQ}
\newblock
\APACjournalVolNumPages{Biometrics}{68}{}{1019--1027}.
\newblock
\begin{APACrefDOI} \doi{10.1111/j.1541-0420.2012.01777.x} \end{APACrefDOI}
\PrintBackRefs{\CurrentBib}

\bibitem [\protect \citeauthoryear {%
Wright%
}{%
Wright%
}{%
{\protect \APACyear {1934}}%
}]{%
Wright1934}
\APACinsertmetastar {%
Wright1934}%
\begin{APACrefauthors}%
Wright, S.%
\end{APACrefauthors}%
\unskip\
\newblock
\APACrefYearMonthDay{1934}{}{}.
\newblock
{\BBOQ}\APACrefatitle {{The method of path coefficients}} {{The method of path
  coefficients}}.{\BBCQ}
\newblock
\APACjournalVolNumPages{The Annals of Mathematical Statistics}{5}{3}{161--215}.
\PrintBackRefs{\CurrentBib}

\bibitem [\protect \citeauthoryear {%
Zheng%
\ \BBA {} van~der Laan%
}{%
Zheng%
\ \BBA {} van~der Laan%
}{%
{\protect \APACyear {2017}}%
}]{%
Zheng2017}
\APACinsertmetastar {%
Zheng2017}%
\begin{APACrefauthors}%
Zheng, W.%
\BCBT {}\ \BBA {} van~der Laan, M.%
\end{APACrefauthors}%
\unskip\
\newblock
\APACrefYearMonthDay{2017}{}{}.
\newblock
{\BBOQ}\APACrefatitle {{Longitudinal Mediation Analysis with Time-varying
  Mediators and Exposures, with Application to Survival Outcomes}}
  {{Longitudinal Mediation Analysis with Time-varying Mediators and Exposures,
  with Application to Survival Outcomes}}.{\BBCQ}
\newblock
\APACjournalVolNumPages{Journal of Causal Inference}{5}{2}{}.
\newblock
\begin{APACrefDOI} \doi{10.1515/jci-2016-0006} \end{APACrefDOI}
\PrintBackRefs{\CurrentBib}

\end{thebibliography}

\newpage


\section*{Supplementary material}

Supplementary material: Clarifying causal mediation analysis for the applied researcher: Defining effects based on what we want to learn

In this supplemental material, we put numerical values into our college prep program example, to give a tangible sense of the natural and interventional (in)direct effects.

First, suppose we are in a setting where both types of effects are identified. (For clarity, recall that identification means the causal contrast can be written as a function of the distribution of the observed variables.) We refer the reader to existing literature, e.g., Imai, Keele, and Tingley (2010), Pearl (2012) and VanderWeele, Vansteelandt, and Robins (2014) for explication of the assumptions required for identification. For our current purpose, one relevant detail to note is that in this case there are no intermediate confounders (see Fig. 1), and the natural and interventional (in)direct effects are equal to each other.

*Figure 1.* The special case with no intermediate confounders

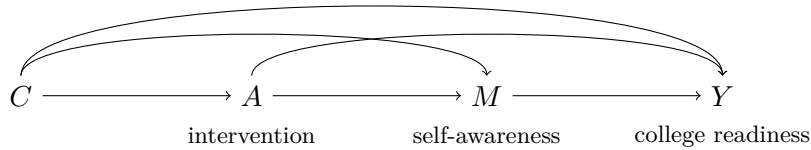

Recall that the mediator (self-awareness) has three possible values, high, medium and low; and the outcome (college readiness) is measured on a 0-to-10 scale. Let's examine a specific $C = c$ subpopulation. Imagine the value $c$ here to mean male high school student in junior year, with a specific GPA, a specific family profile, a specific socio-economic setting, etc., as well as a specific starting level of self-awareness and a specific starting level of college readiness. The idea is that $C$ captures all pre-exposure individual and contextual characteristics that make the individuals exchangeable with respect to the distributions of potential mediators and potential outcomes. This means within this $C = c$ subpopulation, the unconditional distributions of the potential mediators $M(1)$ and $M(0)$ and potential outcomes $Y(1)$ and $Y(0)$ are the same as the distributions of the observed mediator and observed outcome in those with the corresponding exposure conditions; and in addition, the unconditional distribution of the potential outcome $Y(a, m)$ (for a specific $a$ and a specific $m$) is the same as the distribution of the observed outcome in those with exposure $a$ and mediator value $m$. In other words, we get to "observe" the subpopulation distributions of these potential mediators and potential outcomes. We now use these to work out the natural and interventional (in)direct effect for the subpopulation.

The relevant information about these distributions are shown in Table 1. The distributions of potential mediators $M(1)$ and $M(0)$ are shown on the left side of the table. Given two exposure conditions and three possible mediator values, we have six exposure-and-mediator-value combinations. For each combination, there is a distribution of the

Table 1

*Relevant information for identification of natural and interventional (in)direct effects – the special case with no intermediate confounders*

Distributions of potential mediators $M(a')$ within the $C = c$ subpopulation

| potential mediator $M(a')$ | possible value $m$ | chance that potential mediator takes on value $\mathrm{P}(M(a') = m \mid C = c)$ |
|---|---|---|
| $M(0)$ | low | 50% |
| | medium | 40% |
| | high | 10% |
| $M(1)$ | low | 20% |
| | medium | 30% |
| | high | 50% |

Means of potential outcomes $Y(a, m)$ within the $C = c$ subpopulation

| potential outcome $Y(a, m)$ | mean of potential outcome $\mathrm{E}[Y(a, m) \mid C = c]$ |
|---|---|
| $Y(0, \mathrm{low})$ | 3 |
| $Y(0, \mathrm{medium})$ | 5 |
| $Y(0, \mathrm{high})$ | 7 |
| $Y(1, \mathrm{low})$ | 5 |
| $Y(1, \mathrm{medium})$ | 7 |
| $Y(1, \mathrm{high})$ | 8 |

potential outcome, the mean of which is shown on the right side of the table.

In this subpopulation, the mean of $Y(0)$ if 4.2 and the mean of $Y(1)$ is 7.1. This means the subpopulation total effect is

$$\mathrm{TE}_{C=c} = \mathrm{E}[Y(1) \mid C = c] - \mathrm{E}[Y(0) \mid C = c] = 7.1 - 4.2 = 2.9.$$

Looking at the top panel of the table, which concerns exposure condition 0, it is easy to see that 4.2 is simply the weighted average of the three potential outcome means (on the right), where the weights are the probabilities of corresponding potential mediator values (on the left). This weighted average is also the mean of $Y(0, M(0))$ and the mean of $Y(0, \mathcal{M}(0 \mid C))$. The same story applies to 7.1 and the bottom panel of the table. In summary,

$\mathrm{E}[Y(0) \mid C = c] =$

$\mathrm{E}[Y(0, M(0)) \mid C = c] = \mathrm{E}[Y(0, \mathcal{M}(0 \mid C)) \mid C = c] = (50\%)3 + (40\%)5 + (10\%)7 = 4.2,$

$\mathrm{E}[Y(1) \mid C = c] =$

$\mathrm{E}[Y(1, M(1)) \mid C = c] = \mathrm{E}[Y(1, \mathcal{M}(1 \mid C)) \mid C = c] = (20\%)5 + (30\%)7 + (50\%)8 = 7.1.$

The mean of $Y(1, M(0))$ (also the mean of $Y(1, \mathcal{M}(0 \mid C))$) is obtained by combining information from the top-left section (concerning the distribution of $M(0)$) and from the bottom-right section (concerning potential outcomes in conditions with exposure 1) of the table. The mean of $Y(0, M(1))$ (also the mean of $Y(0, \mathcal{M}(1 \mid C))$) is obtained using information from the bottom-left and top-right sections. They are also weighted averages,

$\mathrm{E}[Y(1, M(0)) \mid C = c] = \mathrm{E}[Y(1, \mathcal{M}(0 \mid C)) \mid C = c] = (50\%)5 + (40\%)7 + (10\%)8 = 6.1,$

$\mathrm{E}[Y(0, M(1)) \mid C = c] = \mathrm{E}[Y(0, \mathcal{M}(1 \mid C)) \mid C = c] = (20\%)3 + (30\%)5 + (50\%)7 = 5.6.$

Important note about identification: Here we do not observe any $Y(1, M(0))$ or $Y(0, M(1))$ values. This strategy for estimating their means (by reweighting observed outcomes) is based on a set of conditional independence (also known as ignorabilty, exchangeability or unconfoundedness) assumptions, of which a crucial one is that conditional on pre-exposure covariates $C$, a potential mediator $M(a)$ is independent of potential outcomes $Y(a', m)$, for $a \neq a'$. This assumption amounts to the absence of intermediate confounders.

These four means give us the natural and interventional (in)direct effects for this subpopulation,

$$\text{NDE}_{C=c}(\cdot 0) = \text{IDE}_{C=c}(\cdot 0) = 6.1 - {\color{blue}4.2} = 1.9,$$
$$\text{NDE}_{C=c}(\cdot 1) = \text{IDE}_{C=c}(\cdot 1) = {\color{red}7.1} - 5.6 = 1.5,$$
$$\text{NIE}_{C=c}({\color{blue}0}\cdot) = \text{IIE}_{C=c}({\color{blue}0}\cdot) = 5.6 - {\color{blue}4.2} = 1.4,$$
$$\text{NIE}_{C=c}({\color{red}1}\cdot) = \text{IIE}_{C=c}({\color{red}1}\cdot) = {\color{red}7.1} - 6.1 = 1.0.$$

*Figure 2.* The general case with intermediate confounders ($L$)

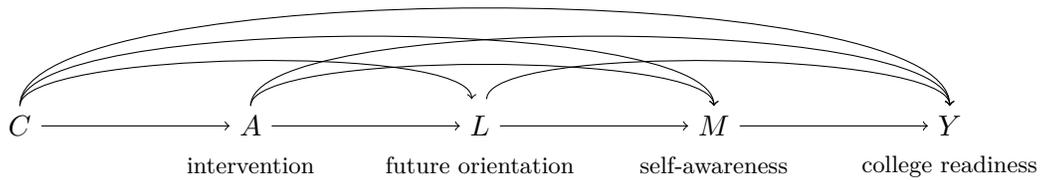

Next, let us turn to the setting where suppose there is an intermediate confounder, *future orientation* (see Fig. 2). For simplicity let this be a binary variable indicating whether the high-schooler demonstrates interest in what they may do for a living and how they may live their life in the future. Table 2 lists information relevant to the current case, which is similar to the information in Table 1, except now the outcome depends on exposure, intermediate confounder, and mediator, so we also require information about the potential distribution of the intermediate confounder under intervention and no intervention.

Table 2
*Relevant information for identification of interventional (in)direct effects – the general case with intermediate confounders*

Distributions of potential confounders $L(a)$ and potential mediators $M(a')$ in the $C = c$ subpopulation

| potential confounder/mediator | possible value | chance of value |
|---|---|---|
| ${\color{blue}L(0)}$ | no | ${\color{blue}70\%}$ |
| | yes | ${\color{blue}30\%}$ |
| | low | ${\color{blue}50\%}$ |
| ${\color{blue}M(0)}$ | medium | ${\color{blue}40\%}$ |
| | high | ${\color{blue}10\%}$ |
| ${\color{red}L(1)}$ | no | ${\color{red}40\%}$ |
| | yes | ${\color{red}60\%}$ |
| | low | ${\color{red}20\%}$ |
| ${\color{red}M(1)}$ | medium | ${\color{red}30\%}$ |
| | high | ${\color{red}50\%}$ |

Means of potential outcomes $Y(a, l, m)$ within the $C = c$ subpopulation

| potential outcome | mean of potential outcome |
|---|---|
| $Y({\color{blue}0}, \text{no}, \text{low})$ | ${\color{blue}2}$ |
| $Y({\color{blue}0}, \text{no}, \text{medium})$ | ${\color{blue}3}$ |
| $Y({\color{blue}0}, \text{no}, \text{high})$ | ${\color{blue}5}$ |
| $Y({\color{blue}0}, \text{yes}, \text{low})$ | ${\color{blue}4}$ |
| $Y({\color{blue}0}, \text{yes}, \text{medium})$ | ${\color{blue}6}$ |
| $Y({\color{blue}0}, \text{yes}, \text{high})$ | ${\color{blue}7}$ |
| $Y({\color{red}1}, \text{no}, \text{low})$ | ${\color{red}3}$ |
| $Y({\color{red}1}, \text{no}, \text{medium})$ | ${\color{red}6}$ |
| $Y({\color{red}1}, \text{no}, \text{high})$ | ${\color{red}7}$ |
| $Y({\color{red}1}, \text{yes}, \text{low})$ | ${\color{red}5}$ |
| $Y({\color{red}1}, \text{yes}, \text{medium})$ | ${\color{red}8}$ |
| $Y({\color{red}1}, \text{yes}, \text{high})$ | ${\color{red}9}$ |

For simplicity, let's focus on only the subpopulation means of $Y({\color{red}1}, {\color{blue}M(0)})$ and of

$Y(1, \mathcal{M}(0 \mid C))$. Note that in the presence of the intermediate confounder, these potential outcomes are the same as $Y(1, L(1), M(0))$ and $Y(1, L(1), \mathcal{M}(0 \mid C))$, respectively.

Since $\mathcal{M}(0 \mid C)$ is randomly drawn from the same distribution for every individual in the $C = c$ subpopulation, it is independent of the naturally occurring $L(1)$ in the subpopulation. This means when we take the weighted average of the subpopulation means of $Y(1, l, m)$ to obtain the subpopulation mean of $Y(1, L(1), \mathcal{M}(0 \mid C))$, the weight is simply the product of the probability that $L(1)$ takes value $l$ and the probability that $M(0)$ takes value $m$.

$$\begin{aligned}
\mathrm{E}[Y(1, L(1), \mathcal{M}(0 \mid C)) \mid C = c] &= (40\%)(50\%)3 + (40\%)(40\%)6 + (40\%)(10\%)7 + \\
&\quad (60\%)(50\%)5 + (60\%)(40\%)8 + (60\%)(10\%)9 \\
&= 5.8.
\end{aligned}$$

Similar weighted averaging obtains the subpopulation means of $Y(1, L(1), \mathcal{M}(1 \mid C))$, $Y(0, L(0), \mathcal{M}(0 \mid C))$, and $Y(0, M(0), \mathcal{M}(1 \mid C))$ to be 7.1, 3.42, and 4.49, respectively. Thus for this subpopulation, we have the interventional (in)direct effects,

$$\begin{aligned}
\mathrm{IDE}_{C=c}(\cdot 0) &= 5.8 - 3.42 = 2.38, \\
\mathrm{IDE}_{C=c}(\cdot 1) &= 7.1 - 4.49 = 2.61, \\
\mathrm{IIE}_{C=c}(0 \cdot) &= 4.49 - 3.42 = 1.07, \\
\mathrm{IIE}_{C=c}(1 \cdot) &= 7.1 - 5.8 = 1.3.
\end{aligned}$$

On the other hand, $M(0)$ is dependent on $L(1)$ in the subpopulation, due to the relationship $L(1) \leftarrow U_L \rightarrow L(0) \rightarrow M(0)$. Therefore, if we want to weight the mean potential outcomes in the bottom-right section of the table to obtain the subpopulation mean of $Y(1, L(1), \mathcal{M}(0 \mid C))$, the weights are no longer the products of pairs of probabilities as above. Instead, the appropriate weights are of the form $\mathrm{P}(L(1) = l, M(0) = m \mid C = c)$, but these probabilities are unidentified because we never observe both $L(1)$ and $M(0)$. The natural (in)direct effects are unidentified in the presence of intermediate confounders.

## References

Imai, K., Keele, L., & Tingley, D. (2010). A general approach to causal mediation analysis. *Psychological Methods*, *15*(4), 309–34. doi: 10.1037/a0020761

Pearl, J. (2012). The causal mediation formula–a guide to the assessment of pathways and mechanisms. *Prevention Science*, *13*(4), 426–36. doi: 10.1007/s11121-011-0270-1

VanderWeele, T. J., Vansteelandt, S., & Robins, J. M. (2014). Effect decomposition in the presence of an exposure-induced mediator-outcome confounder. *Epidemiology*, *25*(2), 300–6. doi: 10.1097/EDE.0000000000000034



\end{document}